\documentclass[onecolumn,showpacs,showkeys,superscriptaddress,pre,12pt]{revtex4-1}
\usepackage[margin=1in]{geometry} 
\usepackage{subcaption}
\usepackage{graphicx}
\usepackage{grffile}
\usepackage{amsmath}
\usepackage{amssymb}
\usepackage{color}
\usepackage{tabularx}
\usepackage{epstopdf}
\usepackage{multirow}
\usepackage{nameref,hyperref}
\usepackage[table]{xcolor}
\usepackage{mathtools}
\usepackage{multirow}
\usepackage{array}
\usepackage{comment}
\usepackage{url}
\usepackage{txfonts}
\usepackage{diagbox}
\usepackage[ddmmyyyy,hhmmss]{datetime}
\usepackage{float}
\usepackage{tabularx}
\usepackage{anyfontsize}
\usepackage[T1]{fontenc}
\usepackage[linesnumbered,ruled,vlined]{algorithm2e}

\newcolumntype{C}{>{\arraybackslash}p{7em}}
\begin{document}
\renewcommand{\thesubsection}{\arabic{subsection}}
\renewcommand{\thesubsubsection}{\arabic{subsection}.\arabic{subsubsection}}
\title{Determining cell population size from cell fraction in cell plasticity models}

\author{Yuman Wang}
\affiliation{School of Mathematical Sciences, Xiamen University, Xiamen 361005, PR China}
\affiliation{National Institute for Data Science in Health and Medicine, Xiamen University, Xiamen 361005, PR China}
\author{Shuli Chen}
\affiliation{School of Mathematical Sciences, Xiamen University, Xiamen 361005, PR China}
\author{Jie Hu}
\email{hujie@xmu.edu.cn}
\affiliation{School of Mathematical Sciences, Xiamen University, Xiamen 361005, PR China}
\author{Da Zhou}
\email{zhouda@xmu.edu.cn}
\affiliation{School of Mathematical Sciences, Xiamen University, Xiamen 361005, PR China}
\affiliation{National Institute for Data Science in Health and Medicine, Xiamen University, Xiamen 361005, PR China}

\begin{abstract}
Quantifying the size of cell populations is crucial for understanding biological processes such as growth, injury repair, and disease progression. Often, experimental data offer information in the form of relative frequencies of distinct cell types, rather than absolute cell counts. This emphasizes the need to devise effective strategies for estimating absolute cell quantities from fraction data. In response to this challenge, we present two computational approaches grounded in stochastic cell population models: the first-order moment method (FOM) and the second-order moment method (SOM). These methods explicitly establish mathematical mappings from cell fraction to cell population size using moment equations of the stochastic models. Notably, our investigation demonstrates that the SOM method obviates the requirement for a priori knowledge of the initial population size, highlighting the utility of incorporating variance details from cell proportions. The robustness of both the FOM and SOM methods was analyzed from different perspectives. Additionally, we extended the application of the FOM and SOM methods to various biological mechanisms within the context of cell plasticity models. Our methodologies not only assist in mitigating the inherent limitations of experimental techniques when only fraction data is available for detecting cell population size, but they also offer new insights into utilizing the stochastic characteristics of cell population dynamics to quantify interactions between different biomasses within the system.
\end{abstract}
\maketitle

\section{Introduction}
\label{}
Both cell proportions and absolute cell numbers are significant in biological research. Cell proportions provide relative abundances of various cell types within tissues or organisms, crucial for understanding tissue structure, function, and developmental processes, as well as studying cell differentiation and changes in cell types under physiological and pathological conditions. Meanwhile, absolute cell numbers offer actual information about the quantity of cells within tissues or organisms, essential for biological processes like growth, injury repair, and disease progression, and serve as important indicators in medical research for disease diagnosis, treatment monitoring, and drug development. In tumor studies, for example, it is essential to consider tumor heterogeneity \cite{2010Tumor}, with one crucial aspect being the variation in the proportions of different types of tumor cells within a population. Additionally, an even more important issue is how the size of the tumor changes, that is, the growth pattern of the tumor \cite{gruber2019growth}. Tumor growth is one of the most significant hallmarks of cancer \cite{hanahan2011hallmarks} and studying tumor growth is of great importance for a comprehensive understanding of tumor occurrence, development, and treatment.

However, in experiments, obtaining dynamic observational data simultaneously for both the relative proportions and absolute quantities of biological systems continuously over time is often quite challenging \cite{yang2012dynamic,overall1973comment,song2000marginal}. Frequently, when measuring a population of multiple cell types, experimental data often present information as relative frequencies of distinct cell types rather than absolute cell quantities \cite{yang2012dynamic,aitchison1982statistical,yuan2019compositional,aitchison2005compositional}. Therefore, an important and intriguing question is whether it is possible, through mathematical and computational means, to infer changes in the absolute quantities of a system only from the knowledge of its component data. 
We address this issue by examining multi-cell type population models. Developing mathematical models has become a crucial approach to gaining a more comprehensive understanding of how various mechanisms impact changes in cell population size, unveiling the dynamic behavior of these populations \cite{spiller2010measurement, paine2016geometrically, stiehl2012mathematical}.
From a mathematical perspective, the relative frequency of a specific cell type is the ratio of the absolute cell count of that cell type to the total population size. For example, consider a scenario with two cell types, $X_t$ and $Y_t$ represent the counts of these two cell types at time $t$, and $N_t = X_t + Y_t$ represents the total population size. Our challenge lies in how to estimate the value of $N_t$ based solely on the provided fraction data, i.e. $X_t/N_t$. It is normally insufficient to resolve $N_t$ from $X_t/N_t$ without access to additional information. Therefore, the key to resolving this issue hinges on the discovery of valuable additional information that can help establish the mapping from $X_t/N_t$ to $N_t$.

In this study, our emphasis is on cell plasticity models—a crucial category of population models involving multiple cell types, characterized by bidirectional transitions between cell states.
Cell plasticity has been reported in many cancers such as breast cancer \cite{chaffer2011normal}, melanoma \cite{quintana2010phenotypic}, and glioblastoma \cite{leder2014mathematical}, where non-stem cancer cells have been observed to reacquire cancer stem cell characteristics through a process of cellular de-differentiation. Considerable modeling efforts have been devoted to studying the impact of cell plasticity, especially in heterogeneous cancer populations \cite{zhou2019invasion, wang2022effect, gupta2011stochastic, jilkine2014effect}.

The cell division pattern is a crucial factor in cell plasticity models. Taking stem cells as an example, they exhibit two distinct division patterns \cite{morrison2006asymmetric}.
The first involves asymmetric cell division \cite{knoblich2008mechanisms,zhong2008neurogenesis}, wherein stem cells undergo an asymmetric division generating two daughter cells with distinct fates: one identical to the mother stem cell and the other differentiating into a non-stem cell state. The second is symmetric cell division \cite{shen2004endothelial}, where stem cells either undergo self-renewal by producing two daughter stem cells identical to the parent, or differentiation by generating two differentiated daughter cells \cite{wu2021stochastic}. 
These two division patterns significantly impact the stochastic fluctuations in the model, thereby influencing our establishment of the mapping relationship from cell proportions to cell population size. One of our major tasks in this study is to explore the effects of different division patterns on establishing this mapping relationship.

We first focus on a cell plasticity model comprising two distinct cell types. We propose a pair of calculation methods for the determination of cell population size based on cell fractions. Specifically, we present the first-order moment method (FOM) and the second-order moment method (SOM) as our primary analytical tools. These methods are instrumental in extracting mathematical mappings from cell fractions to the total cell count.
The FOM approach calls for the availability of prior knowledge regarding the initial population size, serving as a prerequisite for its application. However, the SOM method introduces a paradigm shift, eliminating the necessity for this initial population size. This innovation presents an alternative strategy, particularly valuable in scenarios where the initial population size is elusive or challenging to ascertain. Simultaneously, we have examined the influence of different cell division patterns on the proposed methodology. The results indicate that, for the FOM method, accurate population size calculations can be achieved under both division modes. However, for the SOM method, the accuracy of calculations is higher under symmetric division. Additionally, we consider how cell death affects our mapping establishment. 

Subsequently, we extend the applicability of the FOM and SOM methods to encompass cell plasticity models featuring multiple cell types. This extended framework furnishes explicit expressions for calculating cell population size within the context of multi-cell type systems.
To rigorously validate the robustness and accuracy of our proposed methods, we conducted a series of numerical simulations under various aspects, including different initial values, death rates, division rates, noise interference, and data sample frequencies. 

Our findings not only help alleviate the constraints posed by experimental techniques reliant on fraction data for assessing cell population size but also provide fresh perspectives on utilizing the stochastic aspects of cell population dynamics to measure the interactions among diverse biomasses within the system.

\section{Results}
\label{results}
We propose two methods for determining population size based on cell proportions and apply them to four different cell plasticity models. The accuracy and robustness of these methods were validated using stochastic simulations.

\subsection{Cell plasticity model I: two cell types}
\subsubsection{Model}

\begin{figure}[H]
    \begin{minipage}{0.5\textwidth}  
        \includegraphics[width=0.7\textwidth]{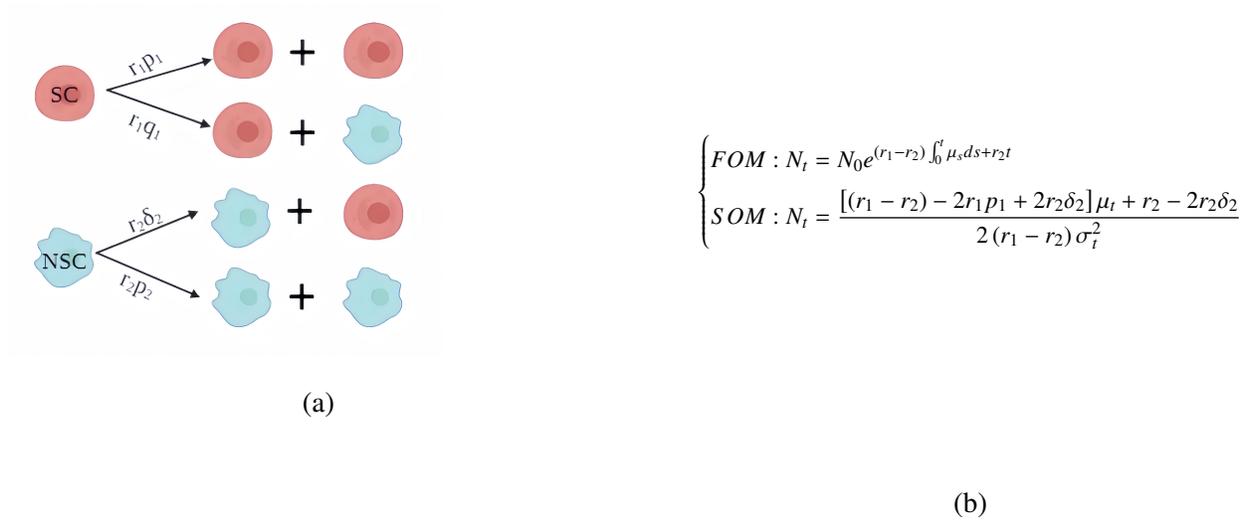} 
        \subcaption{}
    \end{minipage}  
    \hfill  
    \begin{minipage}{0.45\textwidth} 
        \vspace{4\baselineskip}
        \resizebox{\linewidth}{!}{
            $\left\{
            \begin{aligned}
                FOM &:N_{t}=N_{0}e^{(r_{1}-r_{2})\int_{0}^{t}\mu_{s}ds+r_{2}t}\\
                SOM &: N_{t}=\frac{\left[\left(r_{1}-r_{2}\right)-2 r_{1} p_1+2 r_{2} \delta_2\right] \mu_{t}+r_{2}-2 r_{2} \delta_2}{2\left(r_{1}-r_{2}\right) \sigma_{t}^{2}}
            \end{aligned}
            \right.
            $
        }
        \vspace{3.5\baselineskip}
        \subcaption{}
    \end{minipage} 
    \caption{Representation of cell plasticity model I. a) Each compartment represents a specific cell type. Each stem cell can either give birth to two identical stem cells with probability $p_1$ or asymmetric differentiation with probability $q_1$ ($p_1+q_1=1)$. Similar cell division happens to non-stem cells. Due to de-differentiation, each non-stem cell can also give birth to one daughter stem cell and one non-stem cell with probability $\delta_2$. b) The first-order moment (FOM) and second-order moment (SOM) methods for establishing the mappings from cell fraction to cell population size. Here $N_t$ represents the population size at time $t$, $\mu_t$ represents the mean value of SCs proportion, and $\sigma^2_t$ represents the variance of SCs proportion. It shows that FOM needs to use the initial population size, whereas SOM presents an alternative strategy using the variance of SCs fraction.}
    \label{two_asym}
\end{figure}

We first consider a basic compartment model of cell plasticity, consisting of two distinct cell types, as illustrated in Figure \ref{two_asym}(a). The interpretation of these two compartments can vary depending on the biological context. In this instance, we label these two cell types as stem cells (SCs) and non-stem cells (NSCs); however, it is important to note that the model's generality allows it to encompass a wide range of scenarios. To emphasize the plasticity aspect, we assume that cells within these two compartments are capable of inter-conversion, which has been extensively studied in the literature \cite{niu2015phenotypic,zhou2014multi,papaleo2011acidic}. 

We employ a continuous-time Markov chain to capture the population dynamics of the two types of cells. For the stem cell compartment, we assume that each stem cell divides at rate $r_1$, i.e. the waiting time for each cell division event follows an exponential distribution with parameter $r_1$. It can either undergo self-renewal via symmetric division with probability $p_1$ or produce one stem cell and one non-stem cell via asymmetric differentiation with probability $q_1$ ($p_1+q_1=1$).
For the non-stem cell compartment, similarly, each non-stem cell is assumed to divide at rate $r_2$. Either the symmetric division happens with probability $p_2$, or the asymmetric de-differentiation (giving rise to one stem cell and one cell that retains its own identity) happens with probability $\delta_2$ ($p_2+\delta_2=1$). When $\delta_2=0$, indicating the absence of de-differentiation, the model simplifies to the traditional cellular hierarchy model.\\
The schematic representation of cell processes summarizes the above model assumptions as follows ($p_1+q_1=1, p_2+\delta_2=1$):\\
1) $\mathrm{SC}\stackrel{r_1p_1}{\longrightarrow} \mathrm{SC}+\mathrm{SC}.$ \\
2) $\mathrm{SC}\stackrel{r_1q_1}{\longrightarrow} \mathrm{SC}+\mathrm{NSC}.$ \\
3) $\mathrm{NSC}\stackrel{r_2\delta_2}{\longrightarrow} \mathrm{SC}+\mathrm{NSC}.$ \\
4) $\mathrm{NSC}\stackrel{r_2p_2}{\longrightarrow} \mathrm{NSC}+\mathrm{NSC}.$ \\

Let $X_t$ and $Y_t$ denote the number of stem cells and non-stem cells at time $t$, respectively. The stochastic dynamics of the system can be captured by the following master equation \cite{van1992stochastic}:\\
\begin{equation}
\begin{aligned}
\frac{ \partial \varphi_{(i,j)}}{\partial t}
&=\varphi_{(i-1,j)}\cdot (i-1)r_{1}p_1 +\varphi_{(i-1,j)}\cdot jr_{2}\delta_2
\quad+\varphi_{(i,j-1)}\cdot ir_{1}q_1 +\varphi_{(i,j-1)}\cdot (j-1)r_{2}p_2\\
&\quad-\varphi_{(i,j)}\cdot(ir_{1}+jr_{2})\\
\end{aligned}
\label{master}
\end{equation}
where $\varphi_{(i,j)}:=P\left(X_t=i, Y_t=j\right)$, representing the joint probability distribution of $\left(X_t, Y_t\right)$. In what follows
we will use $(X,Y)$ short for $(X_t,Y_t )$. Let $\langle \cdot \rangle$ be the expectation or the moment of a random variable. For example, $\langle X \rangle$ means the first moment of $X$, and $\langle X^2 \rangle$ means the second moment of $X$. We will employ Eq.\eqref{master} to derive moment equations, enabling the utilization of information from moments of cell proportions to deduce the temporal dynamics of population size.

\subsubsection{The first order moment method (FOM)}
\label{2.1.2}
To establish the mapping from cell fraction to the whole population size, we need to make use of the master equation Eq. \eqref{master}  to obtain the moment equations. 
For the first moment, let
$\langle X \rangle=\sum_{i,j}i\varphi_{(i,j)}$,
$\langle Y \rangle=\sum_{i,j}j\varphi_{(i,j)}$, we obtain the ordinary differential equations (ODEs) as follows (refer to Appendix \ref{appA} for details):\\
\begin{equation}
\begin{aligned}
\frac{d\langle X \rangle}{dt}
&=r_1p_1\langle{X}\rangle+r_2\delta_2\langle{Y}\rangle\\
\frac{d\langle Y \rangle}{dt}&=r_1q_1\langle{X}\rangle+r_2p_2\langle{Y}\rangle
\end{aligned}
\end{equation}

Let $R_{X}(t)=X_{t}/N_{t}$ be the cell fraction of stem cells at time $t$, while $R_{Y}(t)=Y_{t}/N_{t}$ be the cell fraction of non-stem cells, where $N_{t}=X_{t}+Y_{t}$ represents the total population size. Let $\mu_t$ and $\gamma_t=1-\mu_{t}$ be the mean values of SC and NSC proportions respectively, i.e.
$\mu_t=\langle{R_{X}(t)}\rangle,\gamma_t=\langle{R_{Y}(t)}\rangle$. We can obtain  the ODEs for the first moment of cell proportions as follows (refer to Appendix \ref{appB} for details):
\begin{equation}
\begin{aligned}
\frac{d \mu_{t}}{d t}=\left(r_{1} p_1-r_{2} \delta_2-\frac{1}{N_{t}} \frac{d N_{t}}{d t}\right) \mu_{t}+r_{2} \delta_2\\
\end{aligned}
\label{mean_A}
\end{equation}

\begin{equation}
\begin{aligned}
\frac{d \gamma_{t}}{d t}=\left(r_{2}p_2-r_{1}q_1-\frac{1}{N_{t}} \frac{d N_{t}}{d t}\right) \gamma_{t}+r_{1}q_1
\end{aligned}
\label{mean_B}
\end{equation}
By the relationship $d \mu_{t}/d t=-d \gamma_{t}/d t$, we have 
\begin{align}
 N_{t}=N_{0}e^{(r_{1}-r_{2})\int_{0}^{t}\mu_{s}ds+r_{2}t}
\label{N_{tmean}}
\end{align}
The integral can be approximated using discrete summation, i.e.\\
\begin{align}
 N_{t_k}=N_{0}e^{(r_{1}-r_{2})\sum_{i=0}^{k-1}\frac{m_{t_i}+m_{t_{i+1}}}{2}\Delta t+r_{2}t_k},
\end{align}
where $t_0=0,~t_{i+1}=t_{i}+\Delta t,~ i=0,1,...,k-1$, $m_{t_i}$ represents the sample point of the mean cell proportion at time $t_i$. This method represents our initial proposal for estimating population size $N_t$ through the utilization of cell proportions, specifically adopting the first-order moment information of stem cell proportion $\mu_t$. As such, we refer to it as the First-Order Moment Method (FOM).

Additionally, it is important to highlight that the size of each subpopulation is calculated by multiplying the total population size by the proportion of each subpopulation. With this understanding of the proportions of each subpopulation, predicting the total population size allows for the prediction of the size of each individual subpopulation.

\subsubsection{The second order moment method (SOM)}
\label{2.1.3}

It is important to highlight that Eq. \eqref{N_{tmean}} relies on the initial state $N_0$. Is it feasible to establish a connection between relative frequency and absolute count without prior knowledge of $N_0$? What specific information regarding cell fractions could help mitigate the impact of the unknown $N_0$? Here we propose an alternative method using the second-order moment of cell fractions.

For the second moment, let $\langle X^2 \rangle=\sum_{i,j}i^2\varphi_{(i,j)}$, 
$\langle Y^2 \rangle=\sum_{i,j}j^2\varphi_{(i,j)}$, based on Eq. \eqref{master} we have the ODEs as follows (refer to  Appendix \ref{appA} for details):\\
\begin{equation}
\frac{d\langle X^{2} \rangle}{dt}=2\langle X^{2} \rangle r_{1}p_1+\langle X \rangle r_{1}p_1
+2\langle XY \rangle r_{2}\delta_2+\langle Y \rangle r_{2}\delta_2\\
\label{sm_X}
\end{equation}
\begin{equation}
\frac{d\langle Y^{2} \rangle}{dt}=2\langle Y^{2} \rangle r_{2}p_2+\langle Y \rangle r_{2}p_2
+2\langle XY \rangle r_{1}q_1+\langle X \rangle r_{1}q_1
\label{sm_Y}
\end{equation}

Let $\sigma_{t}^{2}=Var(R_{X}(t))$ represents the variance of the stem cell proportion. Through straightforward calculations involving Eq.\eqref{mean_A} and Eq.\eqref{sm_X}, we can obtain that:\\
\begin{align}
\frac{d \sigma_{t}^{2}}{d t}=2\left(r_{1} p_1-r_{2} \delta_2-\frac{1}{N_{t}} \frac{d N_{t}}{d t}\right) \sigma_{t}^{2}+\frac{r_{1} p_1-r_{2} \delta_2}{N_{t}} \mu_{t}+\frac{r_{2} \delta_2}{N_{t}}
\end{align}

Likewise,  we obtain the equation for the variance of non-stem cell proportion, i.e. $s_{t}^{2}=Var(R_{Y}(t))$, as follows (See Appendix \ref{appB} for detailed derivation process):\\
\begin{align}
\frac{d s_{t}^{2}}{d t}=2\left[r_{2}p_2-r_{1}q_1-\frac{1}{N_{t}} \frac{d N_{t}}{d t}\right] s_{t}^{2}+\frac{r_{2}p_2-r_{1}p_1}{N_{t}} \gamma_{t}+\frac{r_{1}q_1}{N_{t}}
\end{align}

By using this relation $d \sigma_{t}^{2}/d t=d s_{t}^{2}/d t$, we obtain the second method  for calculating $N_t$ as follows:\\
\begin{equation}
\begin{aligned}
 N_{t}=\frac{\left[\left(r_{1}-r_{2}\right)-2 r_{1} p_1+2 r_{2} \delta_2\right] \mu_{t}+r_{2}-2 r_{2} \delta_2}{2\left(r_{1}-r_{2}\right) \sigma_{t}^{2}}\\
\label{N_{tvar}}
\end{aligned}
\end{equation}

This is the second method we propose for calculating population size based on cell proportions, denoted as the second-order moment method (SOM). From Eq. \eqref{N_{tvar}}, we can see that the total number of cells can be estimated even without knowledge of the initial value $N_0$ provided the variance information of cell fraction. This result is highly inspiring as it indicates that, even in the absence of any information regarding the absolute cell number, it is possible to establish a mapping from cell proportions to cell counts by introducing additional information about random fluctuations in cell fractions. 

We further decompose Eq. \eqref{N_{tvar}}, aiming to gain a more profound understanding of the contributions of different cellular responses to population growth:\\
\begin{equation}
\begin{aligned}
N_{t}&=\underbrace{\frac{(1-2p_1)r_1}{2\left(r_{1}-r_{2}\right) \sigma_{t}^{2}}\mu_{t}}_{SC}+\underbrace{\frac{(1-2\delta_2)r_2}{2\left(r_{1}-r_{2}\right) \sigma_{t}^{2}}\gamma_{t}}_{NSC}\\
&=\underbrace{\frac{\left[\overbrace{p_1}^{symmetric \quad division}-\overbrace{(1-p_1)}^{asymmetric \quad division}\right]r_{1}}{2\left(r_{2}-r_{1}\right) \sigma_{t}^{2}}\mu_{t}}_{SC}+\underbrace{\frac{\left[\overbrace{(1-\delta_2)}^{symmetric \: division}-\overbrace{\delta_2}^{asymmetric \: division}\right]r_2}{2\left(r_{1}-r_{2}\right) \sigma_{t}^{2}}\gamma_{t}}_{NSC}\\
&=\frac{\overbrace{\underbrace{ r_1 p_1\mu_t}_{Self-renewal\: of \:SC}+\underbrace{r_2\delta_2 \gamma_t}_{Asymmetric \:differentiation \:of \:NSC}}^{Net  \:growth  \:rate \:of\:SC}}{2(r_2-r_1)\sigma_{t}^{2}}
+\frac{\overbrace{\underbrace{(1-\delta_2)r_2  \gamma_t}_{Self-renewal \: of \: NSC}+\underbrace{(1-p_1)r_1 \mu_t}_{Asymmetric\:  differentiation \: of \:
 SC}}^{Net \: growth\: rate\: of\: NSC}}{2(r_1-r_2)\sigma_{t}^{2}}
\label{N_{tvar}2}
\end{aligned}
\end{equation}

As depicted above, the denominator represents the interaction between the two cell types, while the numerator can be partitioned into two components. The first component can be regarded as the net growth rate of stem cells, stemming from both the self-renewal of stem cells and the asymmetric de-differentiation of non-stem cells. Similarly, the second component can be expressed as the net growth rate of non-stem cells. Recognizing that the SOM method involves certain approximations that cannot be ignored, we will validate its practicality in the next section through stochastic simulations.

\subsubsection{Methods validation}

\begin{figure}[H]
\centering
\includegraphics[width=0.8\textwidth]{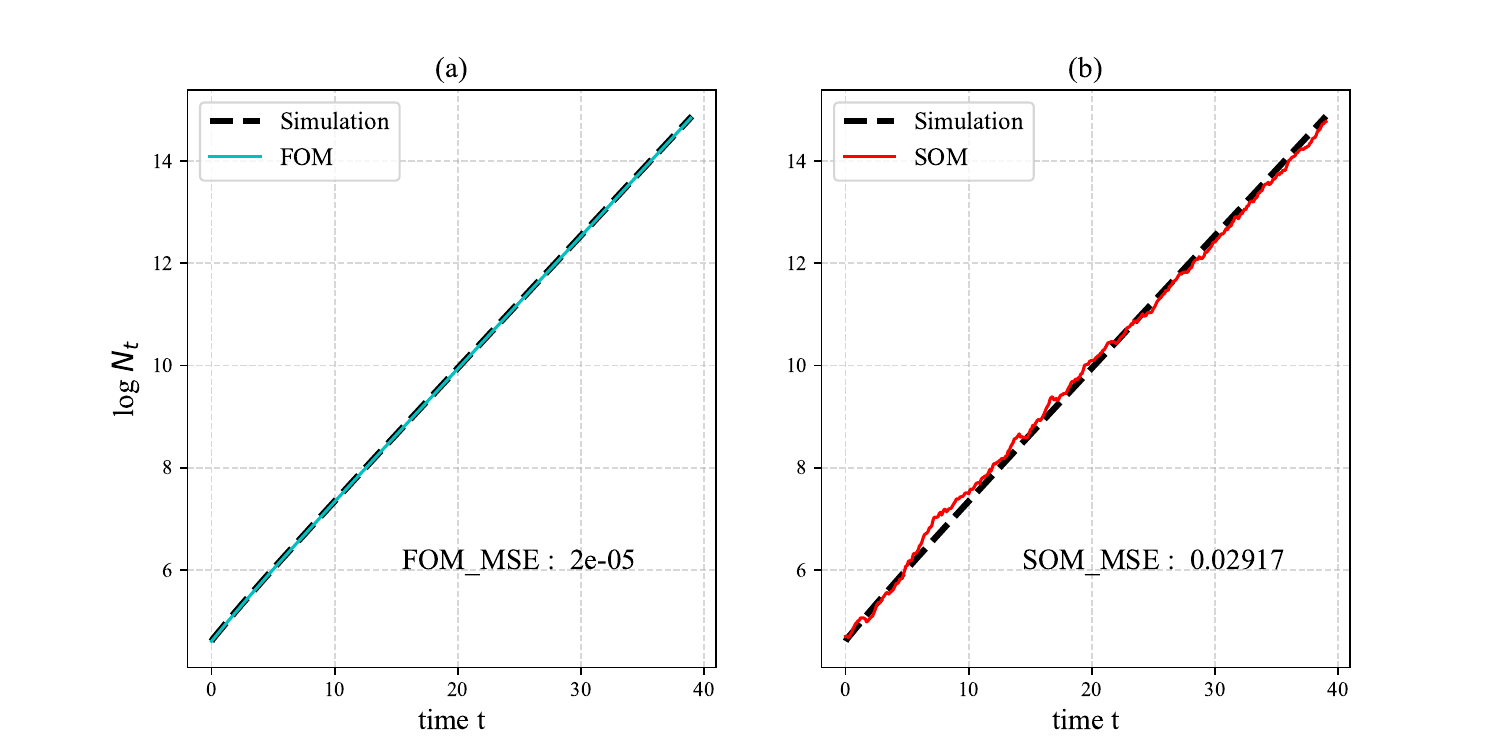}
\caption{Validation of FOM and SOM. In both panels, we depict the logarithmic scale of cell growth, and the black dashed line represents the cell growth obtained from stochastic simulations. $\mathrm{FOM\_MSE}$ denotes the mean squared error between $N_t$ calculated using the FOM method and the simulated $N_t$, whereas $\mathrm{SOM\_MSE}$ represents the equivalent measure for the SOM method. Panel (a) illustrates the comparison between simulation and first-order moment (FOM) method obtained from Eq.\eqref{N_{tmean}} and (b) illustrates the comparison between simulation and second-order moment (SOM) method obtained from Eq.\eqref{N_{tvar}}. The simulations were conducted using data from 100 simulation runs. Parameters: ($r_1,r_2,p_1,\delta_2$)=(0.4,0.2,0.35,0.25).}
\label{fig2}
\end{figure}

\begin{figure}[H]
\centering
\includegraphics[width=0.8\textwidth]{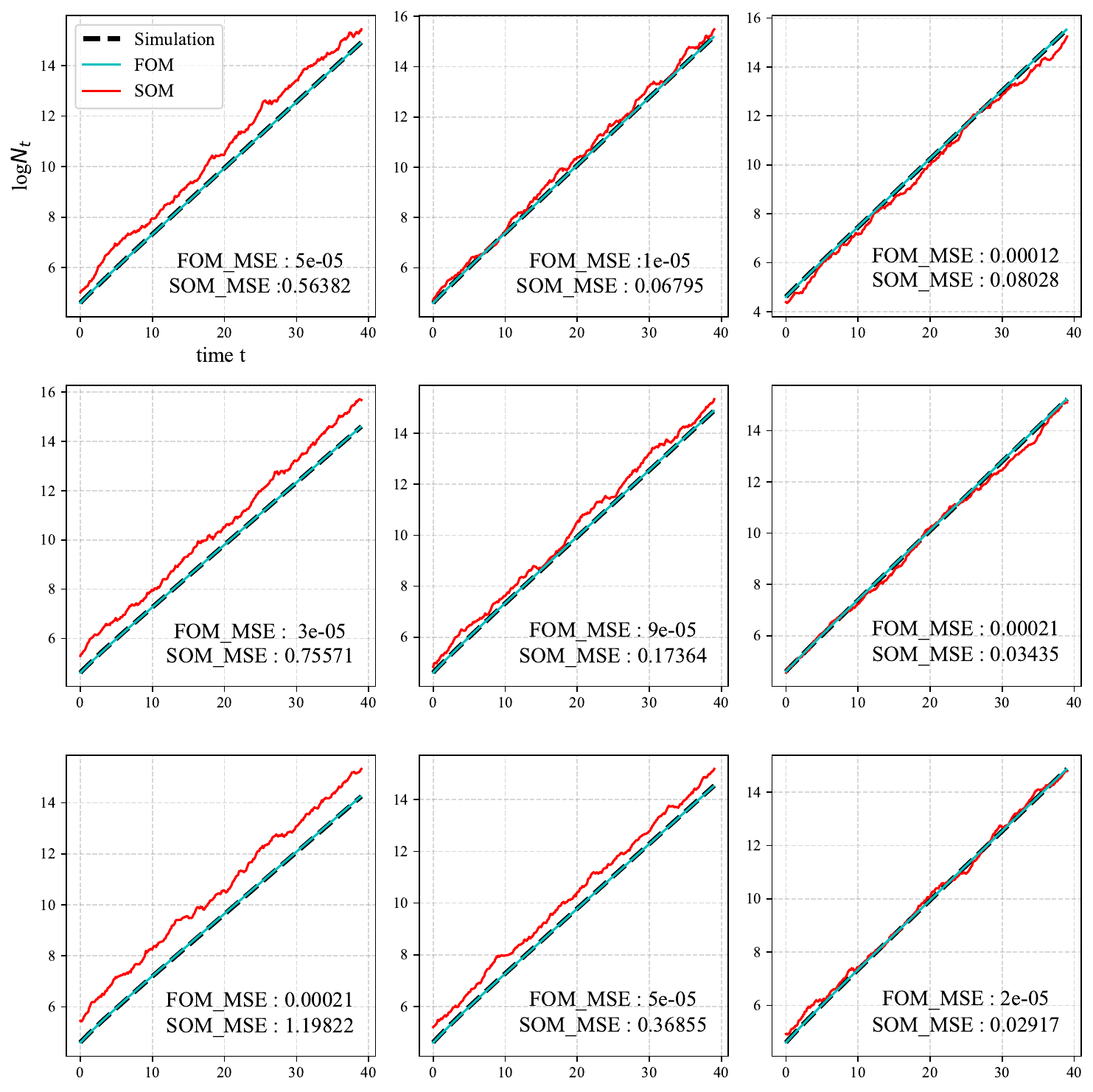}
\caption{Illustration of the robustness of FOM and SOM in cell plasticity model I. The black dashed line represents the growth process using stochastic simulation. The cyan line represents the results obtained from the FOM method, and the red line represents the results obtained from the SOM method. $\mathrm{FOM\_MSE}$ represents the mean squared error between $N_t$ obtained using the FOM method and the simulated $N_t$, while $\mathrm{SOM\_MSE}$ denotes the corresponding result for the SOM method. Parameters: ($r_1,r_2$)=(0.4,0.2), the range of $p_1$ is (0.15, 0.35), the range of $\delta_2$ is (0.25, 0.45), and the number of trajectories is 100.}
\label{two_asym_total}
\end{figure}

To validate FOM and SOM, we conducted individual-based stochastic simulations of cell kinetics using the Gillespie algorithm \cite{Lipshtat2007An,gillespie1977exact}, originally introduced for simulating chemical reactions, allowing us to replicate the random trajectories of the model (see Appendix \ref{appG}). When a cell was randomly chosen for division, it underwent either asymmetric division or self-renewal, determined by their respective probabilities. This approach enables the simulation of changes in population size $N_t$ over time, reflecting the assumptions of the cell plasticity model. Subsequently, we compared the simulation results with those obtained from FOM and SOM methods (Figure \ref{fig2}), employing Mean Squared Error (MSE) as a quantitative statistical analysis metric to assess the proposed methodologies (see Appendix \ref{appG}). The outcomes revealed a commendable consistency between FOM and SOM methods and the simulated results. 
We subsequently assess the robustness of the two methods. As depicted in Fig. \ref{two_asym_total},  we considered the performance of both methods under different reaction rates. The FOM method consistently demonstrates high accuracy, while the SOM method shows reduced accuracy under specific parameter values (see Appendix \ref{appC} for the parameter range discussion). We posit several potential reasons for this disparity. Firstly, the absence of initial value information may introduce larger errors during the estimation process. The significance of initial values in precisely estimating the system state cannot be understated, particularly in the context of the dynamic system. Secondly, the SOM method is more sensitive to parameter values, and the amplification of errors under specific parameter values can adversely affect the accuracy of estimation results. Furthermore, the second-order estimation method introduces certain approximation errors during computation, which can contribute to estimation biases. Nevertheless, the strength of our second-order moment estimation method lies precisely in its independence from initial value information. This characteristic enables us to yield viable estimation results even in situations where accurate initial values are lacking or challenging to obtain. Therefore, both FOM and SOM have their respective strengths and limitations. The FOM exhibits higher accuracy, especially when precise initial values are accessible, rendering it more appropriate under such circumstances. On the other hand, the SOM shines in its ability to operate independently of initial population size, making it more suitable for intricate scenarios with uncertain or unattainable initial states. 

To further assess the effectiveness and robustness of our methods, we conduct a series of simulation experiments under various aspects. Initially, we investigate the impact of different initial values (see \ref{initial condition}). Quantitative analysis is performed for both FOM and SOM across three cases: $N_0$=50, 100, and 200. The findings suggest that while initial values exert some influence on the outcomes, the effect is marginal.

Subsequently, we investigate the impact of death rates(see \ref{death rates}). By comparing scenarios with no mortality events (see Fig. \ref{two_sym_total}), relatively low death rates (see Fig. \ref{two_sym_death_total}), and higher death rates (see Fig. \ref{d=0.05}), we observe that both methods demonstrate resilience to changes in death rates.
Moreover, we examine the impact of additional noise by introducing Gaussian noise of varying intensities (see \ref{noise}). The results indicate that for smaller variance levels, the impact on our methods is negligible. However, as the variance increases, the effectiveness of the SOM method weakens or even fails. This indicates that the SOM method is quite sensitive to noise.
Finally, we explore how the model's performance is affected by measurement frequency by progressively increasing the time intervals and decreasing the sample size from 400 to 80, 40, and eventually 2 (see \ref{time interval}). Our findings indicate that as the sampling frequency decreases, the accuracy of the FOM method diminishes significantly, while the accuracy of the SOM method remains relatively stable. We speculate that this might be attributed to the reduced sampling frequency affecting the approximation of numerical integral values in the FOM method, leading to a deterioration in its performance. In contrast, the SOM method appears to be unaffected by this issue, resulting in minimal changes in its accuracy. This interesting finding suggests that the SOM method can still be utilized even in the presence of missing data.

\subsubsection{Direct transitions}

\begin{figure}[H]
    \centering
    \begin{minipage}{0.5\textwidth} 
        \includegraphics[width=0.6\textwidth]{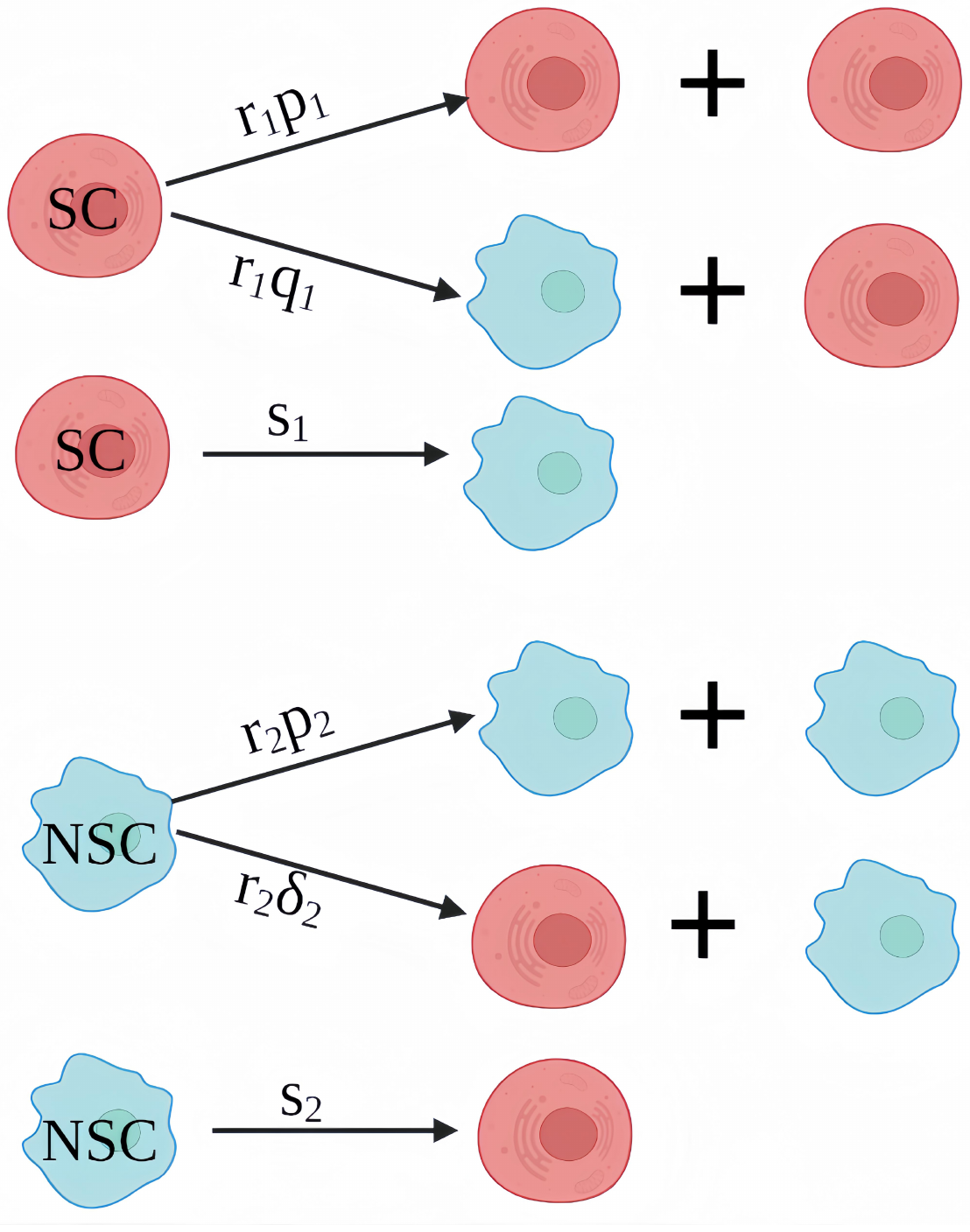} 
        \subcaption{}
    \end{minipage}  
    \hfill  
    \begin{minipage}{0.45\textwidth} 
        \vspace{4\baselineskip} 
        \resizebox{\linewidth}{!}{
            $\left\{
            \begin{aligned}
                FOM &:N_{t}=N_{0}e^{(r_{1}-r_{2})\int_{0}^{t}\mu_{s}ds+r_{2}t}\\
                SOM &: N_{t}=\frac{\left(r_{1}-r_{2}-2r_1q_1+2r_2\delta_2 \right) \mu_{t}+(r_2-2r_{2}\delta_2) }{2\left(r_1-r_2\right) \sigma_{t}^{2}}
            \end{aligned}
            \right.
            $
        }
        \vspace{3\baselineskip} 
        \subcaption{}
    \end{minipage} 
    \caption{Representation of cell plasticity model with direct transitions. (a) Stem cells and non-stem cells undergo direct interconversion at rates $s_1$ and $s_2$, respectively. (b) The first-order moment (FOM) and second-order moment (SOM) methods have been established for mapping from cell proportions to population size.}
    \label{two_trans}
\end{figure}

In the cell plasticity model I (Fig. \ref{two_asym}), we investigated the cellular state transitions induced by cell division but did not consider direct transitions between cellular states \cite{zhou2014multi, gupta2011stochastic}. In this section, we will incorporate direct transitions and assess their impact on our FOM and SOM methods (see Fig.\ref{two_trans}).\\
In addition to the cell transition mechanisms in Model I (Fig. \ref{two_asym}), we add direct transitions between cells. Stem cells (SC) transition directly to non-stem cells (NSC) at a rate of $s_1$, conversely, NSCs can also transition directly to SCs at a rate of $s_2$. The schematic representation of the entire cellular process is summarized as follows ($p_1+q_1=1, p_2+\delta_2=1$):\\
1) $\mathrm{SC}\stackrel{r_1p_1}{\longrightarrow} \mathrm{SC}+\mathrm{SC}.$ \\
2) $\mathrm{SC}\stackrel{r_1q_1}{\longrightarrow} \mathrm{SC}+\mathrm{NSC}.$ \\
3) $\mathrm{SC}\stackrel{s_1}{\longrightarrow} \mathrm{NSC}.$ \\
4) $\mathrm{NSC}\stackrel{r_2\delta_2}{\longrightarrow} \mathrm{SC}+\mathrm{NSC}.$ \\
5) $\mathrm{NSC}\stackrel{r_2p_2}{\longrightarrow} \mathrm{NSC}+\mathrm{NSC}.$ \\
6) $\mathrm{NSC}\stackrel{s_2}{\longrightarrow} \mathrm{SC}.$ 

Following the methodologies outlined in Sec. \ref{2.1.2}and Sec. \ref{2.1.3}, we derive the corresponding FOM (Eq. \ref{FOM_two_trans}) and SOM (Eq. \ref{SOM_two_trans}).\\
\begin{align}
\mathrm{FOM}:N_{t}=N_{0}e^{(r_{1}-r_{2})\int_{0}^{t}\mu_{s}ds+r_{2}t}
\label{FOM_two_trans}
\end{align}

\begin{align}
\mathrm{SOM}:N_{t}=\frac{\left[\left(r_{1}-r_{2}\right)-2 r_1q_1+2 r_2\delta_2\right] \mu_{t}+r_2-2r_2\delta_2}{2\left(r_1-r_2\right) \sigma_{t}^{2}}
\label{SOM_two_trans}
\end{align}

\begin{figure}[H]
\centering
\includegraphics[width=0.8\textwidth]{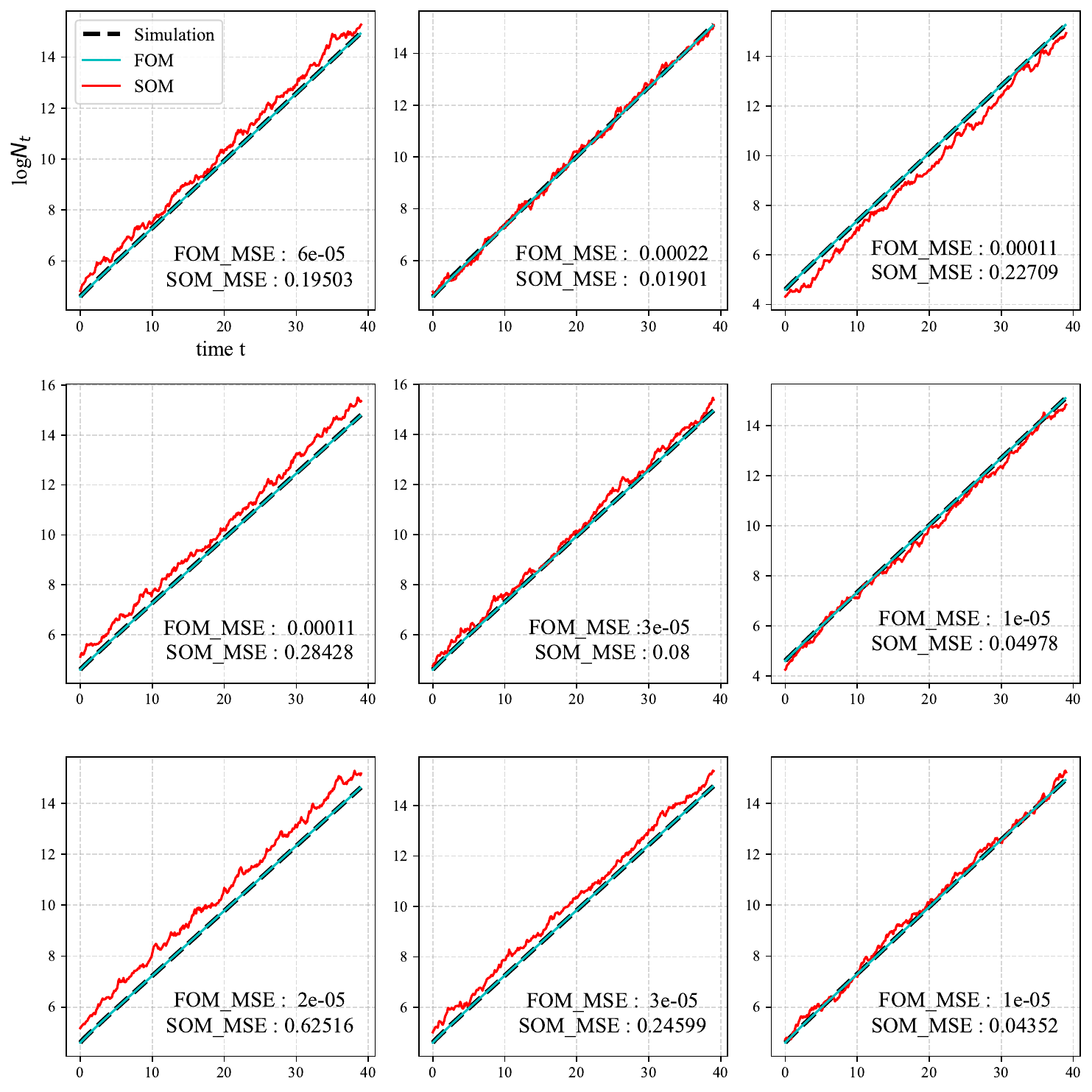}
\caption{Illustration of the robustness of FOM and SOM in cell plasticity model with cell transition. The black dashed line represents the growth process using stochastic simulation. The cyan line represents the results obtained from the FOM method, and the red line represents the results obtained from the SOM method. $\mathrm{FOM\_MSE}$ represents the mean squared error between $N_t$ obtained using the FOM method and the simulated $N_t$, while $\mathrm{SOM\_MSE}$ denotes the corresponding result for the SOM method. Parameters: ($r_1,r_2$)=(0.4,0.2), the range of $p_1$ is (0.15, 0.35), the range of $\delta_2$ is (0.25, 0.45), and the number of trajectories is 100. }
\label{two_asym_trans_total}
\end{figure}
We then validate the two methods using stochastic simulation (Fig.\ref{two_asym_trans_total}). It shows that similar to Fig.\ref{two_asym_total}, the FOM method provides a satisfactory estimation, whereas the SOM method exhibits unignorable errors within specific parameter ranges. However, it is worth noting that the introduction of cell state transitions did not affect the effectiveness of both methods. \\

\subsection{Cell plasticity model II: symmetric division}
\begin{figure}[H]
    \centering
    \begin{minipage}{0.5\textwidth} 
        \includegraphics[width=0.7\textwidth]{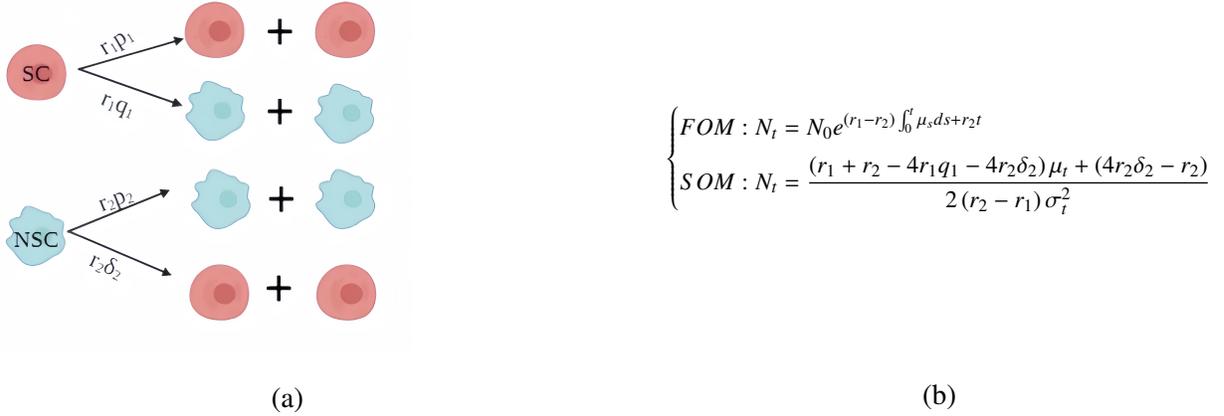} 
        \subcaption{}
    \end{minipage}  
    \hfill  
    \begin{minipage}{0.45\textwidth} 
        \vspace{2\baselineskip} 
        \resizebox{\linewidth}{!}{
            $\left\{
            \begin{aligned}
                FOM &:N_{t}=N_{0}e^{(r_{1}-r_{2})\int_{0}^{t}\mu_{s}ds+r_{2}t}\\
                SOM &: N_{t}=\frac{\left(r_{1}+r_{2}-4r_1q_1-4r_2\delta_2 \right) \mu_{t}+(4r_{2}\delta_2-r_2) }{2\left(r_2-r_1\right) \sigma_{t}^{2}}
            \end{aligned}
            \right.
            $
        }
        \vspace{2.2\baselineskip} 
        \subcaption{}
    \end{minipage} 
    \caption{Representation of cell plasticity model II. (a) Both stem cells and non-stem cells undergo symmetric divisions. (b) Under the symmetric division mode, the first-order moment (FOM) and second-order moment (SOM) methods have been established for mapping from cell proportions to population size.}
    \label{two_sym}
\end{figure}

We have proposed two methods (FOM and SOM) for calculating the total number of cells based on a minimal cell plasticity model composed of two cell types. Previous research has indicated that different organisms may exhibit distinct cell division modes. For instance, asymmetric division has been observed in invertebrates \cite{watt2000out}, while symmetric division is more prevalent in mammals \cite{simons2011strategies}. In this section, we will delve into the derivation and validation of the FOM and SOM methods based on a cell plasticity model characterized by a symmetric division mode (see Fig. \ref{two_sym}). \\
In this model, for each stem cell, division occurs at a rate $r_1$, with a probability $p_1$ of generating two progeny cells identical to the mother cell, or with a probability $q_1$ of producing two identical non-stem cell descendants. Similarly, for each non-stem cell, the division also occurs symmetrically. The schematic representation of cellular processes summarizes the above model assumptions as follows ($p_1+q_1=1,p_2+\delta_2=1)$:\\
1) $\mathrm{SC}\stackrel{r_1p_1}{\longrightarrow} \mathrm{SC}+\mathrm{SC}.$ \\
2) $\mathrm{SC}\stackrel{r_1q_1}{\longrightarrow} \mathrm{NSC}+\mathrm{NSC}.$ \\
3) $\mathrm{NSC}\stackrel{r_2p_2}{\longrightarrow} \mathrm{NSC}+\mathrm{NSC}.$ \\
4) $\mathrm{NSC}\stackrel{r_2\delta_2}{\longrightarrow} \mathrm{SC}+\mathrm{SC}.$ 

Similar to cell plasticity model I, we can derive the master equation for cell plasticity model II, subsequently obtaining the equations for cell quantities. Let $\mu_t$ and $\gamma_t$ represent the mean proportions of the two cell types respectively. The corresponding ODEs are provided below (detailed derivation can be found in Appendix \ref{appD}):\\
\begin{equation}
\begin{aligned}
\frac{d \mu_{t}}{d t}&=\left(r_{1} p_1-r_1q_1-\frac{1}{N_{t}} \frac{d N_{t}}{d t}\right) \mu_{t}+2r_2\delta_2\gamma_{t}\\
\frac{d \gamma_{t}}{d t}&=\left(r_2p_2-r_2\delta_2-\frac{1}{N_{t}} \frac{d N_{t}}{d t}\right) \gamma_{t}+2r_1q_1\mu_t\\
\label{two_sym_mean}
\end{aligned}
\end{equation}
By the relationship $d \mu_{t}/d t=-d \gamma_{t}/d t$, we can obtain the FOM for determining population size from cell proportions:\\
\begin{align}
\mathrm{FOM}:N_{t}=N_{0}e^{(r_{1}-r_{2})\int_{0}^{t}\mu_{s}ds+r_{2}t}
\label{FOM_two_sym}
\end{align}
We denote $\sigma_{t}^{2}, s_{t}^{2}$ as the variances of the proportions of the two cell types, we have (see Appendix \ref{appD} for details):\\
\begin{equation}
\begin{aligned}
\frac{d \sigma_{t}^{2}}{d t}&=2\left(r_1p_1-r_1q_1-2r_2\delta_2-\frac{1}{N_{t}} \frac{d N_{t}}{d t}\right) \sigma_{t}^{2}+\frac{r_{1} -4r_2\delta_2}{N_{t}} \mu_{t}+\frac{4r_2\delta_2}{N_{t}}\\
\frac{d s_{t}^{2}}{d t}&=2\left(r_2p_2-r_2\delta_2-2r_1q_1-\frac{1}{N_{t}} \frac{d N_{t}}{d t}\right) s_{t}^{2}+\frac{r_2-4r_1q_1}{N_{t}} \gamma_{t}+\frac{4r_1q_1}{N_{t}}
\end{aligned}
\label{two_sym_var}
\end{equation}
Based on the relationship $d \sigma_{t}^{2}/dt = ds_{t}^{2}/dt$, we obtain the SOM for calculating $N_t$ as follows:\\
\begin{align}
\mathrm{SOM}:N_{t}=\frac{\left[\left(r_{1}+r_{2}\right)-4 r_1q_1-4 r_2\delta_2\right] \mu_{t}+4r_2\delta_2-r_2}{2\left(r_2-r_1\right) \sigma_{t}^{2}}
\label{SOM_two_sym}
\end{align}

\begin{figure}[H]
\centering
\includegraphics[width=0.8\textwidth]{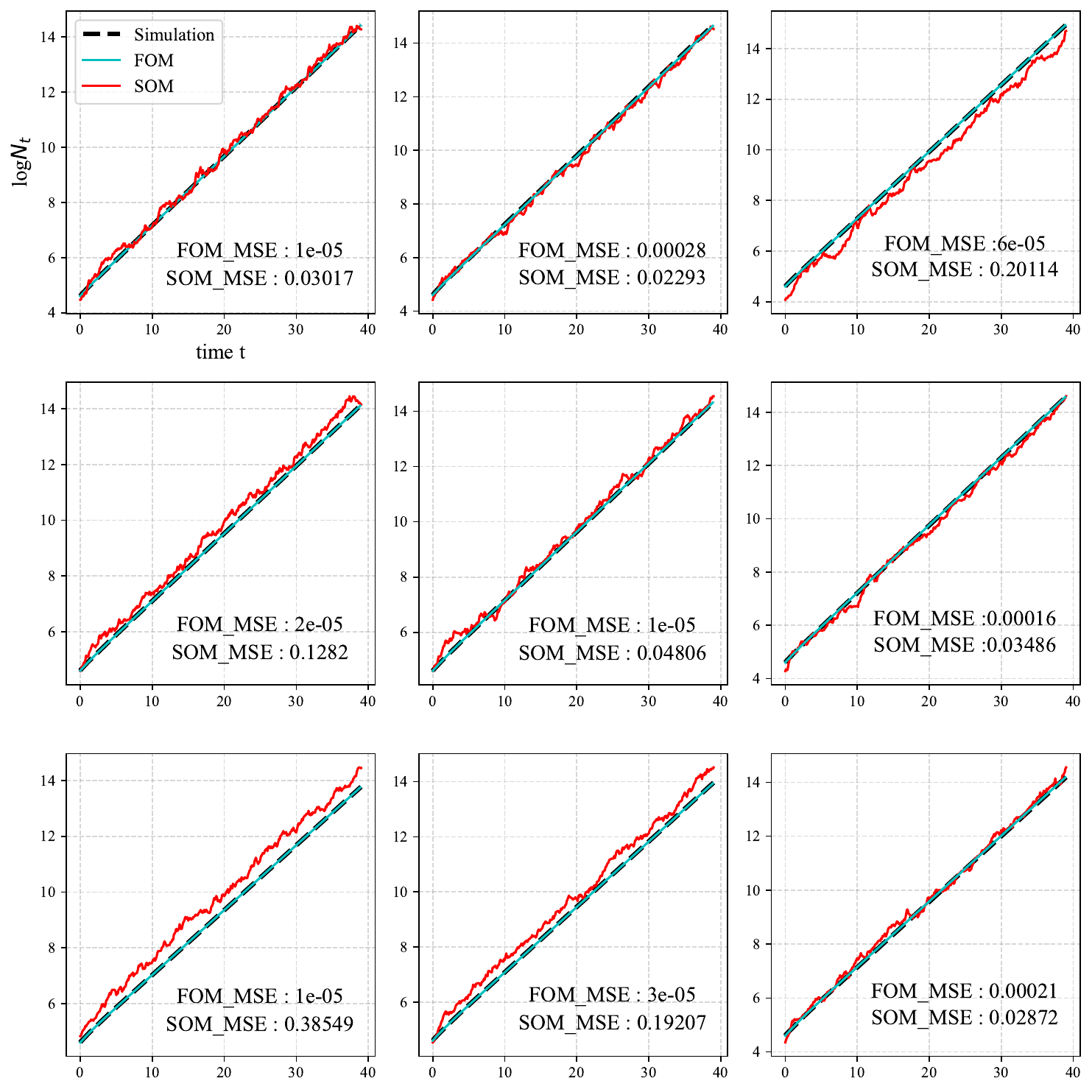}
\caption{Illustration of the robustness of FOM and SOM in cell plasticity model II. The black dashed line represents the growth process using stochastic simulation. The cyan line represents the results obtained from the FOM method, and the red line represents the results obtained from the SOM method. $\mathrm{FOM\_MSE}$ represents the mean squared error between $N_t$ obtained using the FOM method and the simulated $N_t$, while $\mathrm{SOM\_MSE}$ denotes the corresponding result for the SOM method. Parameters: ($r_1,r_2$)=(0.4,0.2), the range of $p_1$ is (0.15, 0.35), the range of $\delta_2$ is (0.25, 0.45), and the number of trajectories is 100. }
\label{two_sym_total}
\end{figure}

We then validate these two methods in cell plasticity model II through stochastic simulation (Fig.\ref{two_sym_total}). It shows that the FOM method performs very well, while the SOM method does not perform as well as the FOM method, similar to the results of the cell plasticity model I shown in Fig. \ref{two_asym_total}. However, it is worth noting that the SOM method performs slightly better in model II compared to model I. This could be because, relative to model I, the symmetric division mode adopted in model II entails a richer set of random fluctuations \cite{wu2021stochastic}. The distinctive feature of the SOM method lies in constructing a mapping from cell proportions to cell population size by supplementing variance information. The richer the random fluctuation information in the model, the more effectively it could provide valuable additional insights.

\subsection{Cell plasticity model III: cell death}
\label{model3}

\begin{figure}[H]
    \begin{minipage}{0.5\textwidth}  
        \includegraphics[width=\textwidth]{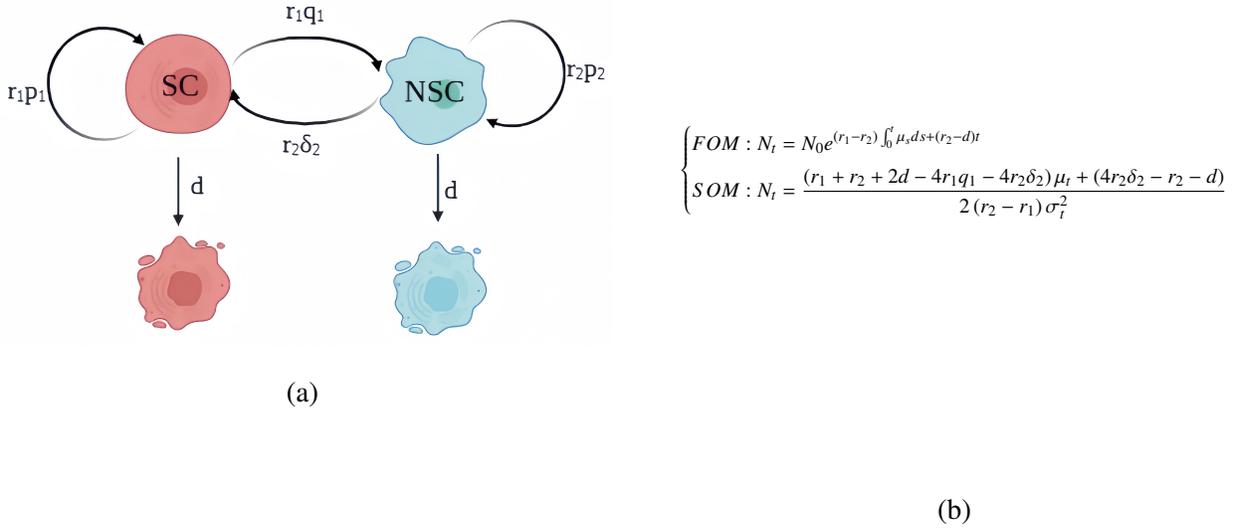} 
        \subcaption{}
    \end{minipage}  
    \hfill  
    \begin{minipage}{0.45\textwidth} 
        \vspace{5\baselineskip}
        \resizebox{\linewidth}{!}{
            $\left\{
            \begin{aligned}
                FOM &:N_{t}=N_{0}e^{(r_{1}-r_{2})\int_{0}^{t}\mu_{s}ds+(r_{2}-d)t}\\
                SOM &: N_{t}=\frac{\left(r_{1}+r_{2}+2d-4r_1q_1-4r_2\delta_2 \right) \mu_{t}+(4r_{2}\delta_2-r_2-d) }{2\left(r_2-r_1\right) \sigma_{t}^{2}}
            \end{aligned}
            \right.
            $
        }
        \vspace{4\baselineskip} 
        \subcaption{}
    \end{minipage} 
    \caption{Representation of cell plasticity model III. a) Each compartment represents a specific cell type. From left to right, it represents stem cells (SCs) and non-stem cells (NSCs) respectively. In addition to self-renewal and differentiation, cells in each compartment also die at a rate of $d$. b) The first-order moment (FOM) and second-order moment (SOM) methods for establishing the mappings from cell fraction to cell population size.}
    \label{two_sym_death}
\end{figure}

We have discussed scenarios involving different cell division modes without considering cell death. Here we introduce cell death in cell plasticity model III (see Fig. \ref{two_sym_death}).\\
Compared to model II, each type of cell undergoes death at a rate of $d$. The schematic representation of cellular processes is given as follows ($p_1+q_1=1, p_2+\delta_2=1$):\\
1) $\mathrm{SC}\stackrel{r_1 p_1}{\longrightarrow} \mathrm{SC}+\mathrm{SC}.$ \\
2) $\mathrm{SC}\stackrel{r_1q_1}{\longrightarrow} \mathrm{NSC}+\mathrm{NSC}.$ \\
3) $\mathrm{SC}\stackrel{d}{\longrightarrow} \varnothing.$ \\
4) $\mathrm{NSC}\stackrel{r_2 \delta_2}{\longrightarrow} \mathrm{SC}+\mathrm{SC}.$ \\
5) $\mathrm{NSC}\stackrel{r_2p_2}{\longrightarrow} \mathrm{NSC}+\mathrm{NSC}.$ \\
6) $\mathrm{NSC}\stackrel{d}{\longrightarrow} \varnothing.$

Let $\mu_t,\gamma_t$ denote the mean values of the proportions of SCs and NSCs, respectively. We have (see Appendix \ref{appE} for more details):\\
\begin{equation}
\begin{aligned}
\frac{d \mu_{t}}{d t}=\left(r_{1} p_1-r_1q_1-d-\frac{1}{N_{t}} \frac{d N_{t}}{d t}\right) \mu_{t}+2r_2\delta_2\gamma_{t}\\
\frac{d \gamma_{t}}{d t}=\left(r_2p_2-r_2\delta_2-d-\frac{1}{N_{t}} \frac{d N_{t}}{d t}\right) \gamma_{t}+2r_1q_1\mu_t\\
\end{aligned}
\end{equation}

Due to the relationship 
$d \mu_{t}/d t=-d \gamma_{t}/d t$, we can derive the FOM for calculating the total cell population size based on cell proportion in this model:\\
\begin{align}
\mathrm{FOM}:N_{t}=N_{0}e^{(r_{1}-r_{2})\int_{0}^{t}\mu_{s}ds+(r_{2}-d)t}
\label{FOM_two_death}
\end{align}

Let $\sigma_{t}^{2}, s_{t}^{2}$ represent the variances of the proportions of SCs and NSCs, respectively. We can obtain that (see Appendix \ref{appE} for details):\\
\begin{align}
\frac{d \sigma_{t}^{2}}{d t}&=2\left(r_1p_1-r_1q_1-d-2r_2\delta_2-\frac{1}{N_{t}} \frac{d N_{t}}{d t}\right) \sigma_{t}^{2}+\frac{r_{1} +d-4r_2\delta_2}{N_{t}} \mu_{t}+\frac{4r_2\delta_2}{N_{t}}\\
\frac{d s_{t}^{2}}{d t}&=2\left(r_2p_2-r_2\delta_2-d-2r_1q_1-\frac{1}{N_{t}} \frac{d N_{t}}{d t}\right) s_{t}^{2}+\frac{r_2+d-4r_1q_1}{N_{t}} \gamma_{t}+\frac{4r_1q_1}{N_{t}}
\end{align}

Based on the relationship $d \sigma_{t}^{2}/dt = ds_{t}^{2}/dt$, we derive the SOM for calculating $N_t$ as follows:\\
\begin{equation}
\mathrm{SOM}:N_{t}=\frac{\left[\left(r_{1}+r_{2}\right)+2d-4 r_1q_1-4 r_2\delta_2\right] \mu_{t}+4r_2\delta_2-r_2-d}{2\left(r_2-r_1\right) \sigma_{t}^{2}}\\
\label{SOM_two_death}
\end{equation}

\begin{figure}[H]
\centering
\includegraphics[width=0.8\textwidth]{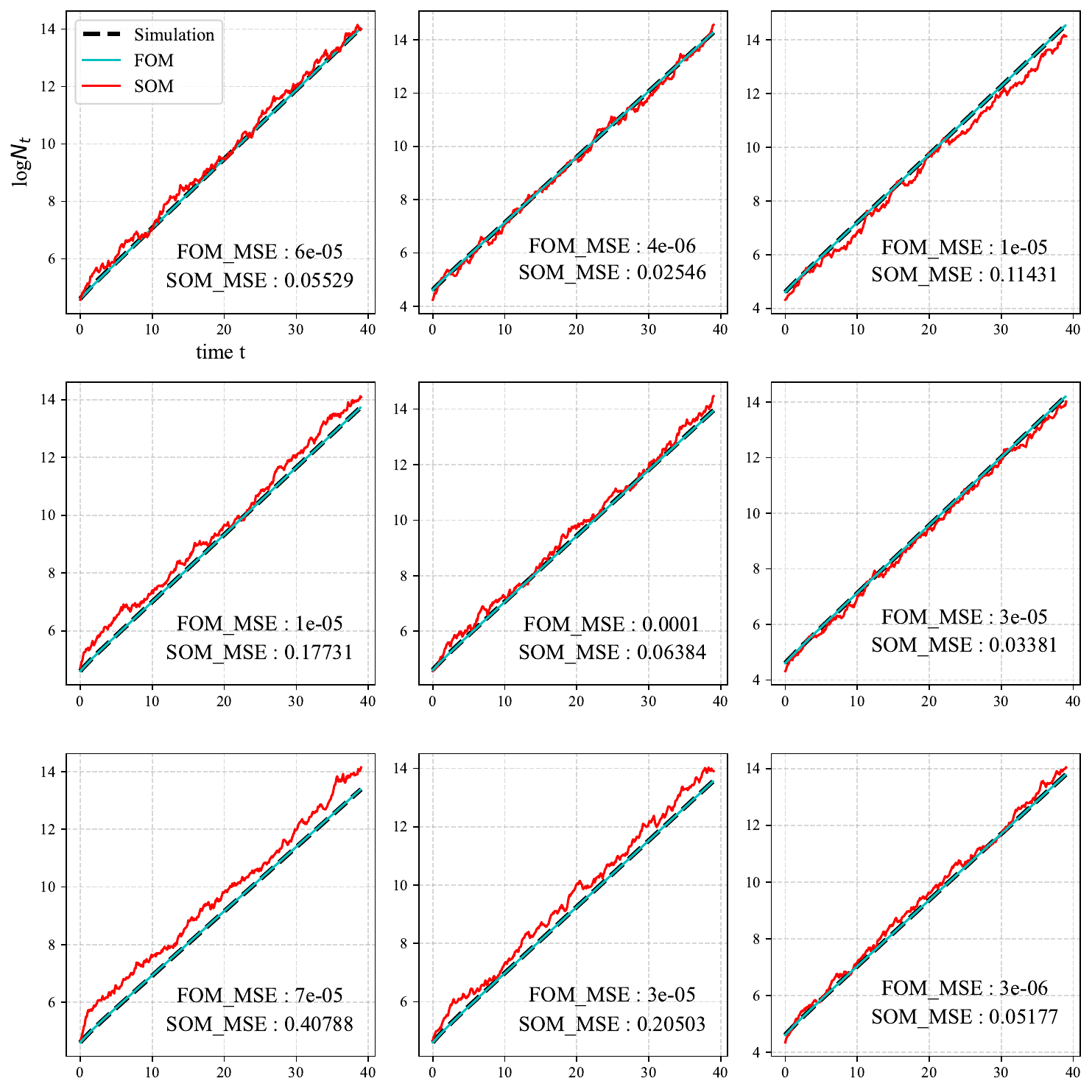}
\caption{Illustration of the robustness of FOM and SOM in cell plasticity model III. The black dashed line in the graph represents the growth process using stochastic simulation. The cyan line represents the results obtained from the FOM method, and the red line represents the results obtained from the SOM method. $\mathrm{FOM\_MSE}$ represents the mean squared error between $N_t$ obtained using the FOM method and the simulated $N_t$, while $\mathrm{SOM\_MSE}$ denotes the corresponding result for the SOM method. Parameters: ($r_1,r_2$)=(0.4,0.2), the range of $p_1$ is (0.15, 0.35), the range of $\delta_2$ is (0.25, 0.45), and the number of trajectories is 100.  }
\label{two_sym_death_total}
\end{figure}
We then validate the two methods using stochastic simulation (Fig.\ref{two_sym_death_total}). It shows that similar to Fig.\ref{two_sym_total}, the FOM method provides a satisfactory estimation, whereas the SOM method exhibits unignorable errors within specific parameter ranges, as shown in the previous two models. However, it is worth noting that the introduction of cell death did not affect the effectiveness of both methods.

\subsection{Cell plasticity model IV: multiple cell types}
\begin{figure}[H]
    \begin{minipage}{0.5\textwidth}  
        \includegraphics[width=\textwidth]{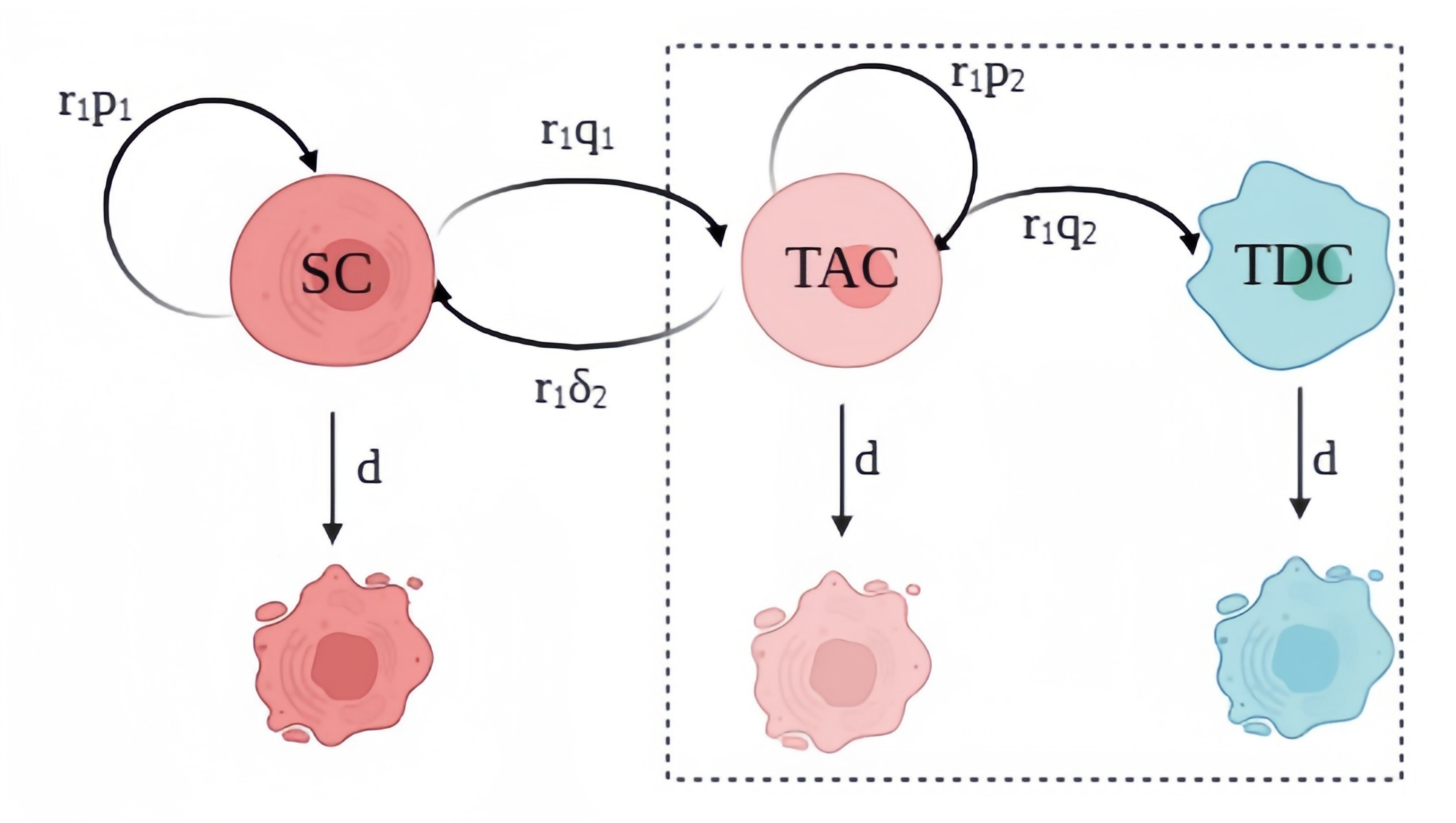} 
        \subcaption{}
    \end{minipage}  
    \hfill  
    \begin{minipage}{0.45\textwidth} 
        \vspace{2\baselineskip}
        \resizebox{\linewidth}{!}{
            $\left\{
            \begin{aligned}
                FOM &:N_{t}=N_{0}e^{(r_{1}-h r_{2})\int_{0}^{t}\mu_{s}ds+(h r_{2}-d)t}\\
                SOM &: N_{t}=\frac{\left[\left(r_{1}+h r_{2}\right)+2d-4 r_1q_1-4h r_2\delta_2\right] \mu_{t}+4h r_2\delta_2-h r_2-d}{2\left(h r_2-r_1\right) \sigma_{t}^{2}}
            \end{aligned}
            \right.
            $
        }
        \vspace{2\baselineskip} 
        \subcaption{}
    \end{minipage} 
    \caption{Representation of cell plasticity model IV. (a) The model comprises three cell types, with each compartment representing a certain level of cellular differentiation. From left to right, they are stem cells (SC), transient amplifying cells (TAC), and terminally differentiated cells (TDC) respectively. b) The first-order moment (FOM) and second-order moment (SOM) methods for establishing the mappings from cell fraction to cell population size. Here $h$ represents the proportion of non-stem cell compartments capable of cell division in the simplified model.}
    \label{three_sym_death}
\end{figure}

In this section, we will extend our methods from two cell types to multiple cell types, taking a three-compartment model as a typical example (see Fig. \ref{three_sym_death}).\\

In this model, each compartment represents a certain level of cellular differentiation (see Fig.\ref{three_sym_death}(a)). 
Specifically, from left to right, they respectively represent stem cells (SCs), transient amplifying cells (TACs), and terminally differentiated cells (TDCs). Each stem cell undergoes division at a rate of $r_1$, followed by symmetric division with a probability of $p_1$ for self-renewal, or with a probability of $q_1$ for symmetric differentiation, generating two TACs. For each TAC, division occurs at a rate of $r_2$, similar to stem cells, involving self-renewal and differentiation. Additionally, due to dedifferentiation, each TAC can generate two stem cells with a probability of $\delta_2$. While TDCs represent mature cells and no longer possess the ability to divide, each TDC only undergoes death at a rate of $d$. The schematic representation of cellular processes summarizes the above model assumptions as follows ($p_1+q_1=1, \delta_2+p_2+q_2=1$):\\
1) $\mathrm{SC}\stackrel{r_1p_1}{\longrightarrow} \mathrm{SC}+\mathrm{SC}.$ \\
2) $\mathrm{SC}\stackrel{r_1q_1}{\longrightarrow} \mathrm{TAC}+\mathrm{TAC}.$ \\
3) $\mathrm{SC}\stackrel{d}{\longrightarrow} \varnothing.$ \\
4) $\mathrm{TAC}\stackrel{r_2\delta_2}{\longrightarrow} \mathrm{SC}+\mathrm{SC}.$ \\
5) $\mathrm{TAC}\stackrel{r_2p_2}{\longrightarrow} \mathrm{TAC}+\mathrm{TAC}.$ \\
6) $\mathrm{TAC}\stackrel{r_2q_2}{\longrightarrow} \mathrm{TDC}+\mathrm{TDC}.$ \\
7) $\mathrm{TAC}\stackrel{d}{\longrightarrow} \varnothing.$\\
8) $\mathrm{TDC}\stackrel{d}{\longrightarrow} \varnothing.$

Compared to the cell plasticity models composed of two cell types, establishing a mapping relationship from cell proportion to cell number in a multi-cell-type model is much more challenging. We here consider consolidating all non-stem cells to form an approximately equivalent two-compartment model (see Fig.\ref{three_sym_death}(a)). Namely, in model IV we consolidate the compartments of TACs and TDCs into a single category termed non-stem cells (NSCs). However, it is important to highlight that in the original model, TDC cells have reached full maturation and no longer possess the ability to undergo cell division. Consequently, in the simplified model, the division capability of NSCs should not be generalized to all cells; rather, it should be attributed to the contribution of TACs, which retain such potential. Here, we employ the proportion of TACs in the entire NSC compartment at steady state to scale the division rate. Let $a, b$, and $c$ be the proportions of SCs, TACs, and TDCs at steady state (determined by the Eq.\ref{de_h} in Appendix \ref{AppF}). We use $h = b/(b+c)$ to represent the proportion of cells capable of division in the simplified model, the division rate of cells in the NSCs compartment is thus given by $r_2 \cdot h$. We validate the effectiveness of this simplification using stochastic simulation (see Fig.\ref{or_si} in Appendix.\ref{AppF}). 

The cellular processes of the simplified model are given as follows:\\
1) $\mathrm{SC} \stackrel{r_{1} p_{1}}{\longrightarrow} \mathrm{SC}+\mathrm{SC}$.\\
2) $\mathrm{SC} \stackrel{r_{1} q_{1}}{\longrightarrow} \mathrm{NSC}+\mathrm{NSC}$.\\
3) $\mathrm{SC} \stackrel{d}{\longrightarrow} \varnothing$.\\
4) $\mathrm{NSC} \stackrel{h\cdot r_{2}(1-\delta_2) }{\longrightarrow} \mathrm{NSC}+\mathrm{NSC}$.\\
5) $\mathrm{NSC} \stackrel{h\cdot r_{2}\delta_2 }{\longrightarrow} \mathrm{SC}+\mathrm{SC}$.\\
6) $\mathrm{NSC} \stackrel{d}{\longrightarrow} \varnothing$.

Let $\mu_t$ and $\gamma_t$ denote the mean proportions of SCs and NSCs, respectively, we have (see Appendix\ref{AppF} for detailed calculations):\\
\begin{equation}
\begin{aligned}
\frac{d \mu_{t}}{d t}=\left(r_{1} p_1-r_1q_1-d-\frac{1}{N_{t}} \frac{d N_{t}}{d t}\right) \mu_{t}+2h\cdot r_2\delta_2\gamma_{t}\\
\frac{d \gamma_{t}}{d t}=\left(h\cdot r_2(1-2\delta_2)-d-\frac{1}{N_{t}} \frac{d N_{t}}{d t}\right) \gamma_{t}+2r_1q_1\mu_t\\
\end{aligned}
\end{equation}

Due to the relationship 
$d \mu_{t}/d t=-d \gamma_{t}/d t$, we can derive the FOM for calculating the total cell population size based on cell proportions:\\
\begin{align}
\mathrm{FOM}:N_{t}=N_{0}e^{(r_{1}-h\cdot r_{2})\int_{0}^{t}\mu_{s}ds+(h\cdot r_{2}-d)t}
\label{FOM_three}
\end{align}

Let $\sigma_{t}^{2}$ and 
$s_{t}^{2}$ be the variances of the proportions of SCs and NSCs, respectively, we have:\\
\begin{align}
\frac{d \sigma_{t}^{2}}{d t}&=2\left(r_1p_1-r_1q_1-d-2h\cdot r_2\delta_2-\frac{1}{N_{t}} \frac{d N_{t}}{d t}\right) \sigma_{t}^{2}+\frac{r_{1} +d-4h\cdot r_2\delta_2}{N_{t}} \mu_{t}+\frac{4h\cdot r_2\delta_2}{N_{t}}\\
\frac{d s_{t}^{2}}{d t}&=2\left(h\cdot r_2(1-2\delta_2)-d-2r_1q_1-\frac{1}{N_{t}} \frac{d N_{t}}{d t}\right) s_{t}^{2}+\frac{h\cdot r_2(1-2\delta_2)+d-4r_1q_1}{N_{t}} \gamma_{t}+\frac{4r_1q_1}{N_{t}}
\end{align}

Based on the relationship $d \sigma_{t}^{2}/dt = ds_{t}^{2}/dt$, we derive the SOM for calculating $N_t$:\\
\begin{equation}
\mathrm{SOM}:N_{t}=\frac{\left[\left(r_{1}+h\cdot r_{2}\right)+2d-4 r_1q_1-4h\cdot r_2\delta_2\right] \mu_{t}+4h\cdot r_2\delta_2-h\cdot r_2-d}{2\left(h\cdot r_2-r_1\right) \sigma_{t}^{2}}\\
\label{SOM_three}
\end{equation}

\begin{figure}[H]
\centering
\includegraphics[width=0.8\textwidth]{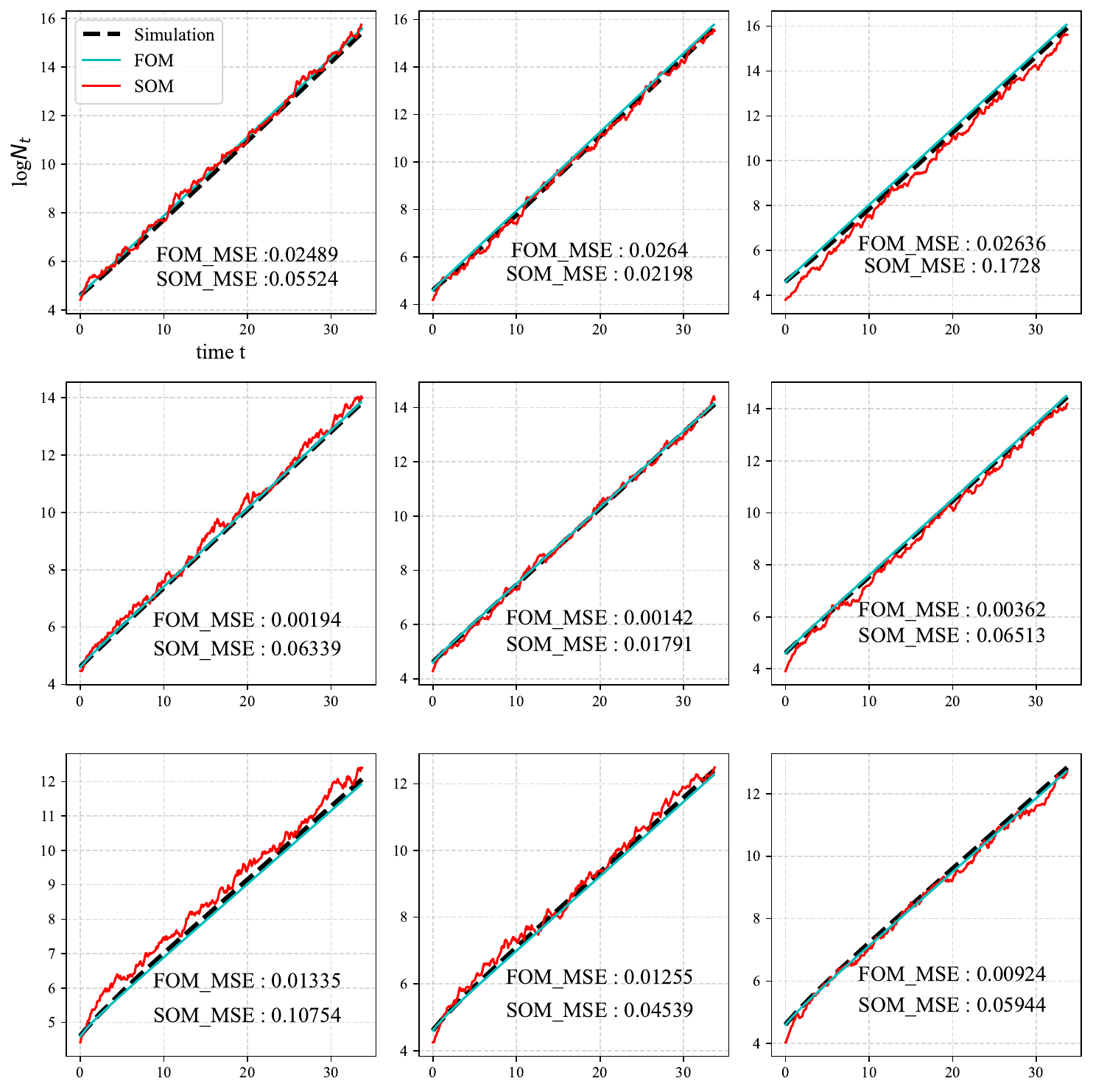}
\caption{llustration of the robustness of FOM and SOM in cell plasticity model IV. The black dashed line signifies the population growth within the original model using stochastic simulation. The cyan line represents the results obtained from the FOM method, and the red line represents the results obtained from the SOM method. $\mathrm{FOM\_MSE}$ represents the mean squared error between $N_t$ obtained using the FOM method and the simulated $N_t$, while $\mathrm{SOM\_MSE}$ denotes the corresponding result for the SOM method. Parameters: ($r_1,r_2$)=(0.5,0.3), the range of $p_1$ is (0.15, 0.35), the range of $\delta_2$ is (0.25, 0.45), and the number of trajectories is 100. }
\label{three_sym_total}
\end{figure}

Fig.\ref{three_sym_total} presents the results of the validation for the FOM and SOM methods in model IV. It shows that our approach successfully extends this capability from the two-compartment model to the three-compartment model, providing a promising avenue for further exploration in more complicated cell plasticity models.

\section{Discussion}

Quantifying cell population sizes holds significant importance in biology. Despite advancements in experimental techniques enabling direct cell counting \cite{robinson2022flow,helleman1969toa,yang2012dynamic}, limitations such as high costs and counting errors persist. Experimental outcomes typically yield relative proportions of cells rather than absolute quantities\cite{yuan2019compositional}. Consequently, formulating effective strategies to determine cell numbers based on cell proportions becomes paramount. This study primarily focuses on a cell plasticity model involving two distinct cell types, for which we have developed a stochastic model to characterize the dynamic behavior of different cell types. We conducted an in-depth exploration of the first-order moment method (FOM) and the second-order moment method (SOM) for determining cell population size based on cell proportions. The application of the FOM method necessitates prior knowledge of initial population size as a prerequisite. Conversely, the SOM method eliminates the need for this prior knowledge using variance information. We have expanded our methodology in various aspects, including different cell division modes, direct transitions between cell states, involvement of cell death processes, and models incorporating multiple compartments. Additionally, we conducted a series of simulation experiments to validate these expansions and simultaneously explored the robustness of the two methods under various aspects, including different initial values, death rates, additional noise, division rates, and sampling frequencies (see Appendix \ref{AppH}).

In the validation results, although the FOM method performs more robustly than the SOM method, we believe that the SOM method provides deeper insights. The FOM method essentially applies average dynamic information, in which the initial cell population size is indispensable when establishing the mapping from cell proportion to cell number. In contrast, the SOM method highlights the importance of random fluctuation information. Interestingly, we found that after incorporating variance information, the SOM method can surprisingly provide a proportion-to-number mapping without knowing the initial cell count. We consider this discovery potentially significant. This is in line with previous studies that variance indeed encapsulates rich biological information and indicates the significant role of variance in cell fate determination  \cite{xue2023logic, chang2008transcriptome, guillemin2020noise}, genetic regulation \cite{jia2017stochastic}, cell signal transduction \cite{levchenko2014cellular,ladbury2012noise}, and other aspects. Incorporating variance into cell proportion modeling contributes to a more comprehensive understanding of key biological features in the dynamic processes of cell populations.

It is noteworthy that, the FOM and SOM methods are grounded in the mean and variance information of cell proportions. To further enhance the accuracy and applicability of the proportion-to-number mapping, high-quality additional information on cell proportion dynamics remains the most crucial aspect. For example, integrating the stochastic trajectory information of cell proportions enables us to more comprehensively understand the dynamic behavior of cell populations. By scrutinizing the time-dependent fluctuations in stochastic trajectories, we can extract more features associated with the evolution of cell populations, thereby facilitating the mapping establishment from cell proportion to cell number. Moreover, rather than employing cell plasticity models with bidirectional cell-state transitions, numerous biological tissues, such as the colon and hematopoietic system, exhibit organization in a hierarchical structure \cite{hou2020noisy, johnston2007mathematical, derenyi2017hierarchical, feliciangeli2022cell, werner2013deterministic}. Examining the adaptation of our methods within a cellular hierarchical framework is also a prospective avenue for our future research.\\

\section*{Acknowledgements}
We appreciate the valuable feedback provided by the two anonymous reviewers.
D.Z. acknowledges the sponsorship by the NSFC (Grant No. 11971405) and the Fundamental Research Funds for the Central Universities in China (Grant No. 20720230023). J.H. acknowledges the sponsorship by the Fundamental Research Funds for the Central Universities in China (Grant No. 20720230024).

\newpage
\appendix
\addcontentsline{toc}{section}{Appendices}
\section*{Appendices}
\numberwithin{equation}{section} 
\renewcommand{\theequation}{\thesection\arabic{equation}}

\section{ Moment equation of cell number}
\label{appA}

Let us consider the first moment $\langle X \rangle$. Noticing that$ \langle X \rangle = \sum_{i,j} i\varphi_{(i, j)}$, we multiply the master equation Eq.\eqref{master} by $i$, and then sum over $i$ and  $j$:
\begin{equation}
\begin{aligned}
\frac{d\langle X \rangle}{dt}
&=\sum_{i,j}i\frac{ \partial  \varphi_{(i,j)}}{\partial t}\\
&=\sum_{i,j}i\varphi_{(i-1,j)}\cdot (i-1)r_{1}p_1 +\sum_{i,j}i\varphi_{(i-1,j)}\cdot jr_{2}\delta_2\quad+\sum_{i,j}i\varphi_{(i,j-1)}\cdot ir_{1}q_1 +\sum_{i,j}i\varphi_{(i,j-1)}\cdot (j-1)r_{2}p_2\\
&\quad-\sum_{i,j}i\varphi_{(i,j)}\cdot(ir_{1}+jr_{2})\\
&=\langle{X^{2}}\rangle r_{1}p_1+\langle{X}\rangle r_{1}p_1+\langle{X^{2}}\rangle r_{1}q_1-\langle{X^{2}}\rangle r_{1}
+\langle{XY}\rangle r_{2}\delta_2+\langle{Y}\rangle r_{2}\delta_2+\langle{XY}\rangle r_{2}p_2-\langle{XY}\rangle r_{2}\\
&=r_{1}p_1\langle{X}\rangle+r_{2}\delta_2\langle{Y}\rangle
\end{aligned}
\end{equation}
Similarly, for $\langle Y \rangle$
\begin{equation}
\begin{aligned}
\frac{d\langle Y \rangle}{dt}=r_{1}q_1\langle{X}\rangle+r_{2}p_2\langle{Y}\rangle
\end{aligned}
\end{equation}
For the second moment $\langle X^2 \rangle$, noticing that$\langle X^2 \rangle=\sum_{i,j}i^2\varphi_{(i,j)}$, and based on Eq.\eqref{master} we have:\\
\begin{equation}
\begin{aligned}
\frac{d\langle X^{2} \rangle}{dt}
&=\sum_{i,j}i^{2}\varphi_{(i-1,j)}\cdot (i-1)r_{1}p_1 +\sum_{i,j}i^{2}\varphi_{(i-1,j)}\cdot jr_{2}\delta_2\quad+\sum_{i,j}i^{2}\varphi_{(i,j-1)}\cdot ir_{1}q_1 +\sum_{i,j}i^{2}\varphi_{(i,j-1)}\cdot (j-1)r_{2}p_2\\
&\quad-\sum_{i,j}i^{2}\varphi_{(i,j)}\cdot(ir_{1}+jr_{2})\\
&=2\langle X^{2} \rangle r_{1}p_1+\langle X \rangle r_{1}p_1
+2\langle XY \rangle r_{2}\delta_2+\langle Y \rangle r_{2}\delta_2
\end{aligned}
\end{equation}
Similarly, for $\langle Y^2 \rangle$ we have:\\
\begin{align}
\frac{d\langle Y^{2} \rangle}{dt}=2\langle Y^{2} \rangle r_{2}p_2+\langle Y \rangle r_{2}p_2
+2\langle XY \rangle r_{1}q_1+\langle X \rangle r_{1}q_1
\end{align}
\newpage
\section{Moment equation of cell proportion}
\label{appB}

Our attention is directed towards elucidating the mapping from cell proportions to population quantities. In the Appendix.\ref{appA}, we have derived the dynamical equations governing cell quantities. Subsequently, we will proceed to deduce the dynamical equations governing cell proportions.  We treat the total number of cells as a deterministic variable \cite{good1972multitype,gillespie2007stochastic}, supported by the observation that $N_t$ exhibits considerably lower fluctuations in comparison to $X_t$ and $Y_t$, and it can be represented by
$N_{t} \approx \langle{X_t}\rangle+\langle{Y_t}\rangle$. $\mu_t$ and $\gamma_t$ are the mean values of the proportions of stem cells and non-stem cells, respectively. Then the equation for the mean cell proportion can be approximated by the following expression:\\
$\begin{aligned}
\centering
\mu_t&=\langle\frac{ X_{t} }{N_t}\rangle \approx \frac{\langle X_{t} \rangle}{N_{t}}\\
& \Longrightarrow \frac{d \langle X_{t}  \rangle}{d t}=\frac{d\left(\mu_{t} N_{t}\right)}{d t}=\mu_{t} \frac{d N_{t}}{d t}+N_{t} \frac{d \mu_{t}}{d t} \\
& \Longrightarrow r_{1} p_1 \langle X_{t}  \rangle+r_{2} \delta_2 \langle Y_{t} \rangle=\mu_{t} \frac{d N_{t}}{d t}+N_{t} \frac{d \mu_{t}}{d t}\\
\end{aligned}$\\
Dividing both sides by $N_{t}$ and rearranging the terms yields the following equation regarding $\mu_{t}$.\\
\begin{equation}
\begin{aligned}
\frac{d \mu_{t}}{d t}=\left(r_{1} p_1-r_{2} \delta_2-\frac{1}{N_{t}} \frac{d N_{t}}{d t}\right) \mu_{t}+r_{2} \delta_2\\
\end{aligned}
\end{equation}
For the non-stem cells, we have:\\
\begin{equation}
\begin{aligned}
\frac{d \gamma_{t}}{d t}=\left(r_{2}p_2-r_{1}q_1-\frac{1}{N_{t}} \frac{d N_{t}}{d t}\right) \gamma_{t}+r_{1}q_1
\end{aligned}
\end{equation}
The variance of stem cell fraction can be expressed as:\\
\begin{equation}
\sigma_{t}^{2}=\operatorname{var}\left(\frac{X_{t}}{X_{t}+Y_{t}}\right) \approx \operatorname{var}\left(\frac{X_{t}}{N_{t}}\right)=\langle\frac{X_{t}^{2}}{N_{t}^{2}}\rangle
-\langle\frac{X_{t}}{N_{t}}\rangle^{2}=\frac{ \langle X_{t}^{2}\rangle}{N_{t}^{2}}-\mu_{t}^{2}
\label{B3}
\end{equation}
By taking the derivative at both ends of Eq.\ref{B3}, we can get the ODE of $\sigma_t^{2}$ as follows:\\
\begin{equation}
\frac{d \sigma_{t}^{2}}{d t}=2\left(r_{1} p_1-r_{2} \delta_2-\frac{1}{N_{t}} \frac{d N_{t}}{d t}\right) \sigma_{t}^{2}+\frac{r_{1} p_1-r_{2} \delta_2}{N_{t}} \mu_{t}+\frac{r_{2} \delta_2}{N_{t}}
\end{equation}
We obtain the variance of the non-stem cell proportion in the same way, as outlined below:\\
\begin{align}
\frac{d s_{t}^{2}}{d t}=2\left[r_{2}p_2-r_{1}q_1-\frac{1}{N_{t}} \frac{d N_{t}}{d t}\right] s_{t}^{2}+\frac{r_{2}p_2-r_{1}p_1}{N_{t}} \gamma_{t}+\frac{r_{1}q_1}{N_{t}}
\end{align}

\newpage
\section{Applicability of SOM}
\label{appC}

Due to the mathematical approximations involved in the derivation of SOM, it is necessary to discuss its applicability. It is challenging to obtain sufficient and necessary conditions for its applicability, so here we provide the necessary conditions for the applicability of the SOM method via steady-state analysis. Noticing that
\begin{equation}
\begin{aligned}
\frac{d \mu_{t}}{d t}&=\left(r_{1} p_1-r_{2} \delta_2-\frac{1}{N_{t}} \frac{d N_{t}}{d t}\right) \mu_{t}+r_{2} \delta_2\\
\frac{1}{N_{t}} \frac{d N_{t}}{d t}&=\left(r_{1}-r_{2}\right)\mu_{t}+r_{2}\\
&\Longrightarrow -(r_1-r_2)\mu_{t}^{2}+[r_1p_1-r_2(1+\delta_2)]\mu_t+r_2\delta_2=0\\
&\Longrightarrow \mu_{\infty} = \frac{-[r_1p_1-r_2(1+\delta_2)] -\sqrt{\Delta}}{-2(r_1-r_2)}\\
\text{where} \quad \Delta &= [r_1p_1-r_2(1+\delta_2)]^{2}+4(r_1-r_2)r_2\delta_2
\label{SSX}
\end{aligned}
\end{equation}
For the non-stem cell proportion, we have $\gamma_{\infty}=1-\mu_{\infty}$.
That is, $\mu_{\infty}$ and $\gamma_{\infty}$ are the proportions of stem cells and non-stem cells at steady state respectively. Note that the population size should be positive, We conduct discussions on parameter scopes for different scenarios:\\
(i) For $r_1>r_2$:\\
\begin{equation*}
N_{t}=\frac{\left[\left(r_{1}-r_{2}\right)-2 r_{1} p_1+2r_{2} \delta_2\right] \mu_{\infty}+r_{2}-2 r_{2} \delta_2}{2\left(r_{1}-r_{2}\right) \sigma_{t}^{2}} > 0
\Longrightarrow r_1(1-2p_1)\mu_{\infty}+r_2(1-2\delta_2)\gamma_{\infty}>0.\\  
\end{equation*}
When $0<p_1<0.5$ and $0<\delta_2<0.5$, we can ensure the establishment of the aforementioned relationship.\\
(ii) For $r_1 < r_2$:\\
\begin{equation*}
N_{t}=\frac{\left[\left(r_{1}-r_{2}\right)-2 r_{1} p_1+2 r_{2} \delta_2\right] \mu_{\infty}+r_{2}-2 r_{2} \delta_2}{2\left(r_{1}-r_{2}\right) \sigma_{t}^{2}} > 0
\Longrightarrow r_1(1-2p_1)\mu_{\infty}+r_2(1-2\delta_2)\gamma_{\infty}<0
\end{equation*}
When $0.5<p_1<1, and 0.5<\delta_2<1$, we can ensure the establishment of the aforementioned relationship.\\

\newpage
\section{Derivation of cell plasticity model II}
\label{appD}
Through model II, our objective is to thoroughly investigate the influence of symmetric division patterns on the FOM method and the SOM method. Based on specific cellular responses, we can obtain the master equation that reflects the population dynamics as follows:\\
\begin{equation}
\begin{aligned}
\frac{ \partial \varphi_{(i,j)}}{\partial t}
&=\varphi_{(i-1,j)}\cdot (i-1)r_{1}p_1 +\varphi_{(i+1,j-2)}\cdot (i+1)r_1q_1+ \varphi_{(i-2,j+1)}\cdot (j+1)r_2\delta_2+\varphi_{(i,j-1)}\cdot (j-1)r_2p_2\\
&-\varphi_{(i,j)}(ir_1p_1+ir_1q_1+jr_2p_2+jr_2\delta_2)
\end{aligned}
\label{master_sym}
\end{equation}

Based on Eq.\ref{master_sym}, we can obtain the differential equations for the first and second moments of cells as follows:\\
\begin{equation}
\left\{
\begin{aligned}
&\frac{d\langle X\rangle}{d t}=\langle X\rangle r_{1} p_{1}-\langle X\rangle r_{1} q_{1}+2\langle Y\rangle r_2\delta_2 \\
&\frac{d\langle Y\rangle}{d t}=2\langle X\rangle r_{1} q_{1}+\langle Y\rangle r_{2} p_{2}-\langle Y\rangle r_2\delta_2 \\
\end{aligned}
\label{D2}
\right.
\end{equation}

\begin{equation}
\left\{
\begin{aligned}
&\frac{d\langle X^2\rangle}{d t}=\langle X^2\rangle (2r_{1} p_{1}-2 r_{1} q_{1})+r_1\langle X\rangle+4\langle XY\rangle r_2\delta_2+4\langle Y\rangle r_2\delta_2\\
&\frac{d\langle Y^2\rangle}{d t}=4\langle X\rangle r_{1} q_{1}+4\langle XY\rangle r_{1} q_{1}+\langle Y^2\rangle (2r_{2} p_{2}-2 r_{2} \delta_{2} )+r_2\langle Y\rangle\\
\end{aligned}
\label{D3}
\right.
\end{equation}

Similar to the approach in Appendix\ref{appB}, we can obtain the following differential equations for the mean of cell proportions:\\
\begin{equation}
\left\{
\begin{aligned}
\mathrm{SC}:&\frac{d \mu_t}{d t} = \left(r_1p_1-r_1q_1-d-\frac{1}{N_t} \frac{d N_t}{d t}\right)\mu_t+2r_2\delta_2\gamma_t \\
\mathrm{NSC}:&\frac{d \gamma_t}{d t} = \left(r_2p_2-r_2\delta_2-\frac{1}{N_t} \frac{d N_t}{d t}\right)\gamma_t +2r_1q_1 \mu_t\\
\end{aligned}
\label{D4}
\right.
\end{equation}
\\

For the variance of cell proportions, we have:
\begin{equation}
\left\{
\begin{aligned}
\mathrm{SC}: \frac{d \sigma_{t}^{2}}{d t}&=2\left(r_1p_1-r_1q_1-2r_2\delta_2-\frac{1}{N_{t}} \frac{d N_{t}}{d t}\right) \sigma_{t}^{2}+\frac{r_1-4r_2\delta_2}{N_{t}} \mu_t+\frac{4r_2\delta_2}{N_t}\\
\mathrm{NSC}: \frac{d s_t^{2}}{d t}&=2\left(r_2p_2-r_2q_2-2r_1q_1-\frac{1}{N_{t}} \frac{d N_{t}}{d t}\right) s_t^{2}+\frac{r_2-4r_1q_1}{N_{t}} \gamma_t+4\frac{r_1q_1}{N_{t}}\\
\end{aligned}
\label{D5}
\right.
\end{equation}

\newpage
\section{Derivation of cell plasticity model III}
\label{appE}
In model III, we incorporate a plasticity model for cell death to explore the impact of the death mechanism on the FOM and SOM methods. From the specific responses of cells in the system, we can obtain the master equation that reflects the population dynamics as follows:\\
\begin{equation}
\begin{aligned}
\frac{ \partial \varphi_{(i,j)}}{\partial t}
&=\varphi_{(i-1,j)}\cdot (i-1)r_{1}p_1 +\varphi_{(i+1,j-2)}\cdot (i+1)r_1q_1+\varphi_{(i+1,j)}\cdot (i+1)d+ \varphi_{(i-2,j+1)}\cdot (j+1)r_2\delta_2\\
&+\varphi_{(i,j-1)}\cdot (j-1)r_2p_2+\varphi_{(i,j+1)}\cdot d-\varphi_{(i,j)}(ir_1p_1+ir_1q_1+jr_2p_2+jr_2\delta_2+
(i+j)d)
\end{aligned}
\label{master_sym_death}
\end{equation}

According to Eq.\ref{master_sym_death}, we can obtain the differential equations for the first and second moments of cells as follows:\\
\begin{equation}
\left\{
\begin{aligned}
&\frac{d\langle X\rangle}{d t}=\langle X\rangle r_{1} p_{1}-\langle X\rangle r_{1} q_{1}-\langle X\rangle d+2\langle Y\rangle r_2\delta_2 \\
&\frac{d\langle Y\rangle}{d t}=2\langle X\rangle r_{1} q_{1}+\langle Y\rangle r_{2} p_{2}-\langle Y\rangle r_2\delta_2 -d\langle Y\rangle\\
\end{aligned}
\label{E2}
\right.
\end{equation}

\begin{equation}
\left\{
\begin{aligned}
&\frac{d\langle X^2\rangle}{d t}=\langle X^2\rangle (2r_{1} p_{1}-2 r_{1} q_{1}-2d)+(r_1+d)\langle X\rangle+4\langle XY\rangle r_2\delta_2+4\langle Y\rangle r_2\delta_2\\
&\frac{d\langle Y^2\rangle}{d t}=4\langle X\rangle r_{1} q_{1}+4\langle XY\rangle r_{1} q_{1}+\langle Y^2\rangle (2r_{2} p_{2}-2 r_{2} \delta_{2}-2d )+(r_2+d)\langle Y\rangle\\
\end{aligned}
\label{E3}
\right.
\end{equation}

Similar to the approach in Appendix\ref{appB}, we can obtain the following differential equations for the mean of cell proportions:\\
\begin{equation}
\left\{
\begin{aligned}
&\frac{d \mu_t}{d t} = \left(r_1p_1-r_1q_1-d-\frac{1}{N_t} \frac{d N_t}{d t}\right)\mu_t+2r_2\delta_2\gamma_t \\
&\frac{d \gamma_t}{d t} = \left(r_2p_2-r_2\delta_2-d-\frac{1}{N_t} \frac{d N_t}{d t}\right)\gamma_t +2r_1q_1 \mu_t\\
\end{aligned}
\label{E4}
\right.
\end{equation}

For the variance of cell proportions, we have:\\
\begin{equation}
\left\{
\begin{aligned}
\mathrm{SC}: \frac{d \sigma_{t}^{2}}{d t}&=2\left(r_1p_1-r_1q_1-2r_2\delta_2-d-\frac{1}{N_{t}} \frac{d N_{t}}{d t}\right) \sigma_{t}^{2}+\frac{r_1+d-4r_2\delta_2}{N_{t}} \mu_t+\frac{4r_2\delta_2}{N_t}\\
\mathrm{NSC}: \frac{d s_t^{2}}{d t}&=2\left(r_2p_2-r_2q_2-d-2r_1q_1-\frac{1}{N_{t}} \frac{d N_{t}}{d t}\right) s_t^{2}+\frac{r_2+d-4r_1q_1}{N_{t}} \gamma_t+4\frac{r_1q_1}{N_{t}}\\
\end{aligned}
\label{E5}
\right.
\end{equation}

\newpage
\section{Derivation of cell plasticity model IV}
\label{AppF}
In model IV, we extend the FOM method and the SOM method to a model with three compartments. Based on the specific cellular responses outlined in the main text, we can obtain the temporal evolution of cell states and the chemical master equation.\\
\begin{equation}
\begin{aligned}
\frac{ \partial \varphi_{(i,j,k)}}{\partial t}
&=\varphi_{(i-1,j,k)}\cdot (i-1)r_{1}p_1 +\varphi_{(i+1,j-2,k)}\cdot (i+1)r_1q_1+ \varphi_{(i+1,j,k)}\cdot(i+1)d+\varphi_{(i-2,j+1,k)}\cdot (j+1)r_2\delta_2\\
&+\varphi_{(i,j-1,k)}\cdot (j-1)r_2p_2+\varphi_{(i,j+1,k-2)}\cdot (j+1)r_2q_2+\varphi_{(i,j+1,k)}\cdot (j+1)d+\varphi_{(i,j,k+1)}\cdot (k+1)d\\
&-\varphi_{(i,j,k)}(ir_1p_1+ir_1q_1+id+jr_2p_2+jr_2\delta_2+jr_2q_2+jd+kd)
\end{aligned}
\label{master_three}
\end{equation}

We calculated the proportion of TAC cells to the total number of TAC and TDC cells at a steady state, enabling an appropriate scaling transformation for the division rate in the non-stem cell compartment in the simplified model. To achieve this, we obtain the first-order moment equation for cell quantities based on the master equation (Eq.\ref{master_three}). \\

\begin{equation}
\left\{
\begin{aligned}
&\frac{d\langle X\rangle}{d t}=\langle X\rangle r_{1} p_{1}-\langle X\rangle r_{1} q_{1}-\langle X\rangle d + 2r_2\delta_2 \langle Y\rangle\\
&\frac{d\langle Y\rangle}{d t}=2\langle X\rangle r_{1} q_{1}+\langle Y\rangle r_{2} p_{2}-\langle Y\rangle r_{2} q_{2}-\langle Y\rangle d -r_2\delta_2 \langle Y\rangle\\
&\frac{d\langle Z\rangle}{d t}=2\langle Y\rangle r_{2} q_{2}-\langle Z\rangle d \\
\end{aligned}
\right.
\end{equation}

Define $A_t:=\langle\frac{X_t}{N_t}\rangle$,$B_t:=\langle\frac{Y_t}{N_t}\rangle$,$C_t:=\langle\frac{Z_t}{N_t}\rangle$, then we can obtain the differential equation for the mean cell ratio:\\
\begin{equation}
\left\{
\begin{aligned}
&\frac{d A_t}{d t} = \left(r_1p_1-r_1q_1-d-\frac{1}{N_t} \frac{d N_t}{d t}\right)A_t +2r_2\delta_2 B_t\\
&\frac{d B_t}{d t} = \left(r_2p_2-r_2q_2-d-r_2\delta_2-\frac{1}{N_t} \frac{d N_t}{d t}\right)B_t +2r_1q_1 A_t\\
&\frac{d C_t}{d t} = \left(-d-\frac{1}{N_t} \frac{d N_t}{d t}\right)C_t +2r_2q_2 B_t\\
&\frac{1}{N_t} \frac{d N_t}{d t}=r_1A_t+r_2B_t-d
\label{F3}
\end{aligned}
\right.
\end{equation}

Let $A_{\infty}=a, B_{\infty}=b,$ and $C_{\infty}=c$ denote the values of the three cell proportions at steady state, respectively. Setting the right-hand side of the Eq.\ref{F3} to zero, and after a straightforward simplification, it can be expressed as follows: \\
\begin{equation}
\left\{
\begin{aligned}
& \left(r_1p_1-r_1q_1\right)a-r_1a^{2}-r_2ab+2r_2\delta_2b =0\\
&\left(r_2p_2-r_2q_2-r_2\delta_2\right)b -r_1ab-r_2b^{2}+2r_1q_1a=0\\
&\left(-r_1a-r_2b \right)c+2r_2q_2b=0\\
\end{aligned}
\label{de_h}
\right.
\end{equation}

The proportion $h=b/(b+c)$ of cell division occurring in the compartment of non-stem cells can be determined by Eq.\ref{de_h}. 
We employ simulation algorithms to validate the models before and after simplification.
\begin{figure}[H]
\centering
\includegraphics[width=0.5\textwidth]{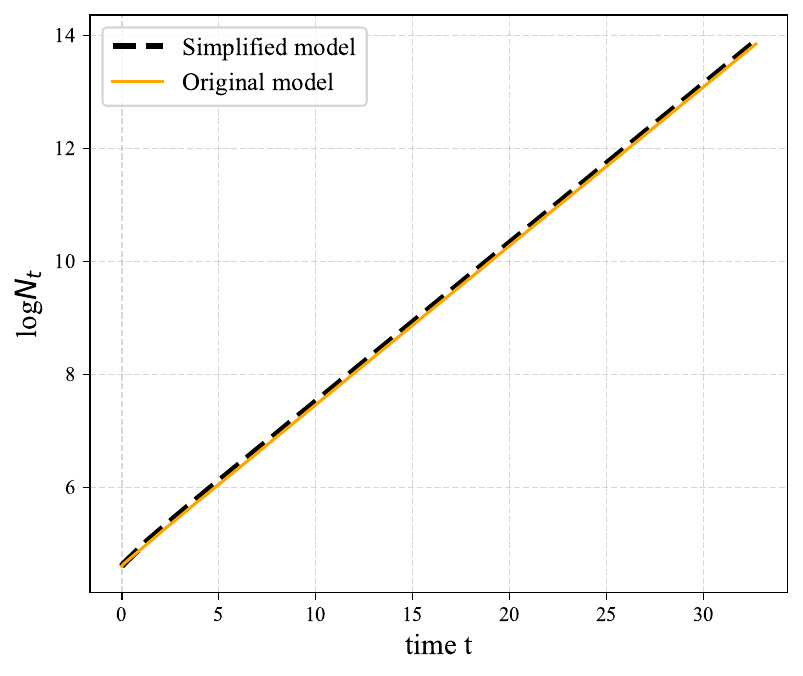}
\caption{Validation of the simplified model. The dashed black line represents the simplified model, while the orange solid line represents the original three-compartment model. Parameters:($r_1,r_2,p_1,p_2,\delta_2$)=(0.5,0.3,0.25,0.5,0.35). }
\label{or_si}
\end{figure}
In Fig.\ref{or_si}, the black dashed line represents the quantity of a specific population in the simplified model, while the orange solid line represents the quantity of the population in the original model. The good fit between the two suggests that the approximation method we employed for simplifying the model has a negligible impact on the estimation of the population quantity of interest, hence validating its reasonability.

Then the dynamical behavior of the simplified model can be characterized by the following master equation:\\
\begin{equation}
\begin{aligned}
\frac{ \partial \varphi_{(i,j)}}{\partial t}
&=\varphi_{(i-1,j)}\cdot (i-1)r_{1}p_1 +\varphi_{(i+1,j-2)}\cdot (i+1)r_1q_1+\varphi_{(i+1,j)}\cdot (i+1)d\\
&+ \varphi_{(i-2,j+1)}\cdot (j+1)h\cdot r_2\delta_2+\varphi_{(i,j-1)}\cdot (j-1)h\cdot r_2p_2+\varphi_{(i,j+1)}\cdot d\\
&-\varphi_{(i,j)}(ir_1p_1+ir_1q_1+jh\cdot r_2p_2+jh\cdot r_2\delta_2+
(i+j)d)
\end{aligned}
\label{master_three_sym}
\end{equation}

According to Eq.\ref{master_three_sym}, we can obtain the differential equations for the first and second moments of cells as follows:\\
\begin{equation}
\left\{
\begin{aligned}
&\frac{d\langle X\rangle}{d t}=\langle X\rangle r_{1} p_{1}-\langle X\rangle r_{1} q_{1}-\langle X\rangle d+2\langle Y\rangle h\cdot r_2\delta_2 \\
&\frac{d\langle Y\rangle}{d t}=2\langle X\rangle r_{1} q_{1}+\langle Y\rangle h\cdot r_{2} p_{2}-\langle Y\rangle h\cdot r_2\delta_2 -d\langle Y\rangle\\
\end{aligned}
\label{F5}
\right.
\end{equation}

\begin{equation}
\left\{
\begin{aligned}
&\frac{d\langle X^2\rangle}{d t}=\langle X^2\rangle (2r_{1} p_{1}-2 r_{1} q_{1}-2d)+(r_1+d)\langle X\rangle+4\langle XY\rangle h\cdot r_2\delta_2+4\langle Y\rangle h\cdot r_2\delta_2\\
&\frac{d\langle Y^2\rangle}{d t}=4\langle X\rangle r_{1} q_{1}+4\langle XY\rangle r_{1} q_{1}+\langle Y^2\rangle (2h\cdot r_{2} p_{2}-2 h\cdot r_{2} \delta_{2}-2d )+(h\cdot r_2+d)\langle Y\rangle\\
\end{aligned}
\label{F6}
\right.
\end{equation}

Similar to the approach in Appendix\ref{appB}, we can obtain the following differential equations for the mean of cell proportions:\\
\begin{equation}
\left\{
\begin{aligned}
&\frac{d \mu_t}{d t} = \left(r_1p_1-r_1q_1-d-\frac{1}{N_t} \frac{d N_t}{d t}\right)\mu_t+2h\cdot r_2\delta_2\gamma_t \\
&\frac{d \gamma_t}{d t} = \left(h\cdot r_2p_2-h\cdot r_2\delta_2-d-\frac{1}{N_t} \frac{d N_t}{d t}\right)\gamma_t +2r_1q_1 \mu_t\\
\end{aligned}
\label{F7}
\right.
\end{equation}

For the variance of cell proportions, we have:\\
\begin{equation}
\left\{
\begin{aligned}
SC: \frac{d \sigma_{t}^{2}}{d t}&=2\left(r_1p_1-r_1q_1-2h\cdot r_2\delta_2-d-\frac{1}{N_{t}} \frac{d N_{t}}{d t}\right) \sigma_{t}^{2}+\frac{r_1+d-4h\cdot r_2\delta_2}{N_{t}} \mu_t+\frac{4h\cdot r_2\delta_2}{N_t}\\
NSC: \frac{d s_t^{2}}{d t}&=2\left(h\cdot r_2p_2-h\cdot r_2q_2-d-2r_1q_1-\frac{1}{N_{t}} \frac{d N_{t}}{d t}\right) s_t^{2}+\frac{h\cdot r_2+d-4r_1q_1}{N_{t}} \gamma_t+4\frac{r_1q_1}{N_{t}}\\
\end{aligned}
\label{F8}
\right.
\end{equation}

\newpage
\section{Stochastic simulation and Statistical analysis}
\label{appG}
\subsection{Stochastic simulation}
In this study, we utilized the Gillespie algorithm \ref{Gillespie} to generate simulated data reflecting the dynamic behavior of cell populations. This algorithm selects the subsequent cellular response based on the individual cell reaction rates and subsequently updates population data and temporal information. 

\begin{algorithm}[H]
\KwData{Initial time $t = 0$ and initial state $S$ with initial cell populations}
\KwResult{Simulation until time $T_{\text{max}}$ is reached}
initialization\;
\While{$t < T_{\text{max}}$}{
Calculate all possible cell reaction rates based on the current state $S$\;
Compute the total reaction rate $R_{\text{total}} = \sum(rate_i)$, where $rate_i$ is the rate of the $i$-th cell reaction\;
Generate random numbers $u$ and $v$ such that $0 < u < 1$ and $0 < v < 1$\;
Calculate the time step for the next reaction, $dt = -(1/R_{\text{total}}) * \ln(u)$\;
Compare $v * R_{\text{total}}$ with the cumulative reaction rates to determine which cell reaction occurs\;
Update time $t = t + dt$\;
Update the state $S$\;
}
\caption{Gillespie Algorithm for Cell Dynamics}
\label{Gillespie}
\end{algorithm}

\subsection{Statistical analysis}
We selected Mean Squared Error (MSE) as a quantitative statistical analysis metric to assess the fit between our proposed methods and simulated data, quantitatively evaluating its accuracy and reliability in predicting the dynamic behavior of cell populations.\\
\begin{align}
\text{MSE} = \frac{1}{n} \sum_{i=1}^{n} (y_i - \hat{y}_{i})^{2}.
\end{align}
Here, $n$ represents the sample size, $y_{i}$ denotes the observed values (simulated generated $N_t$), and $\hat{y}_{i}$ stands for the corresponding predicted values (obtained from the FOM or SOM method for $N_t$). We display the Mean Squared Error (MSE) of both methods in the results figures.

Moreover, we conducted a quantitative analysis to evaluate the robustness of our methodology, including variations in initial conditions ($N_0$=50, 100, 200, Appendix \ref{initial condition}), different mortality rates (d=0.01, d=0.05, see Appendix \ref{death rates}), and diverse rates of cellular division (various combinations of $r_ip_i$, where parameters are selected according to Appendix \ref{appC}). Additionally, we implemented sparse sampling of simulated data at different time intervals (Appendix \ref{time interval}) and introduced different Gaussian noise (Appendix \ref{noise}) to validate the resilience of our approach.

\newpage
\section{Robustness analysis of FOM and SOM methods under various aspects}
\label{AppH}

\subsection{Different initial conditions}
\label{initial condition}
In this section, we focus on the robustness of FOM and SOM methods under different initial values $N_0$. We consider three different levels of $N_0$, namely 50, 100, and 200. \\

\begin{figure}[H]
\centering
\includegraphics[width=0.8\textwidth]{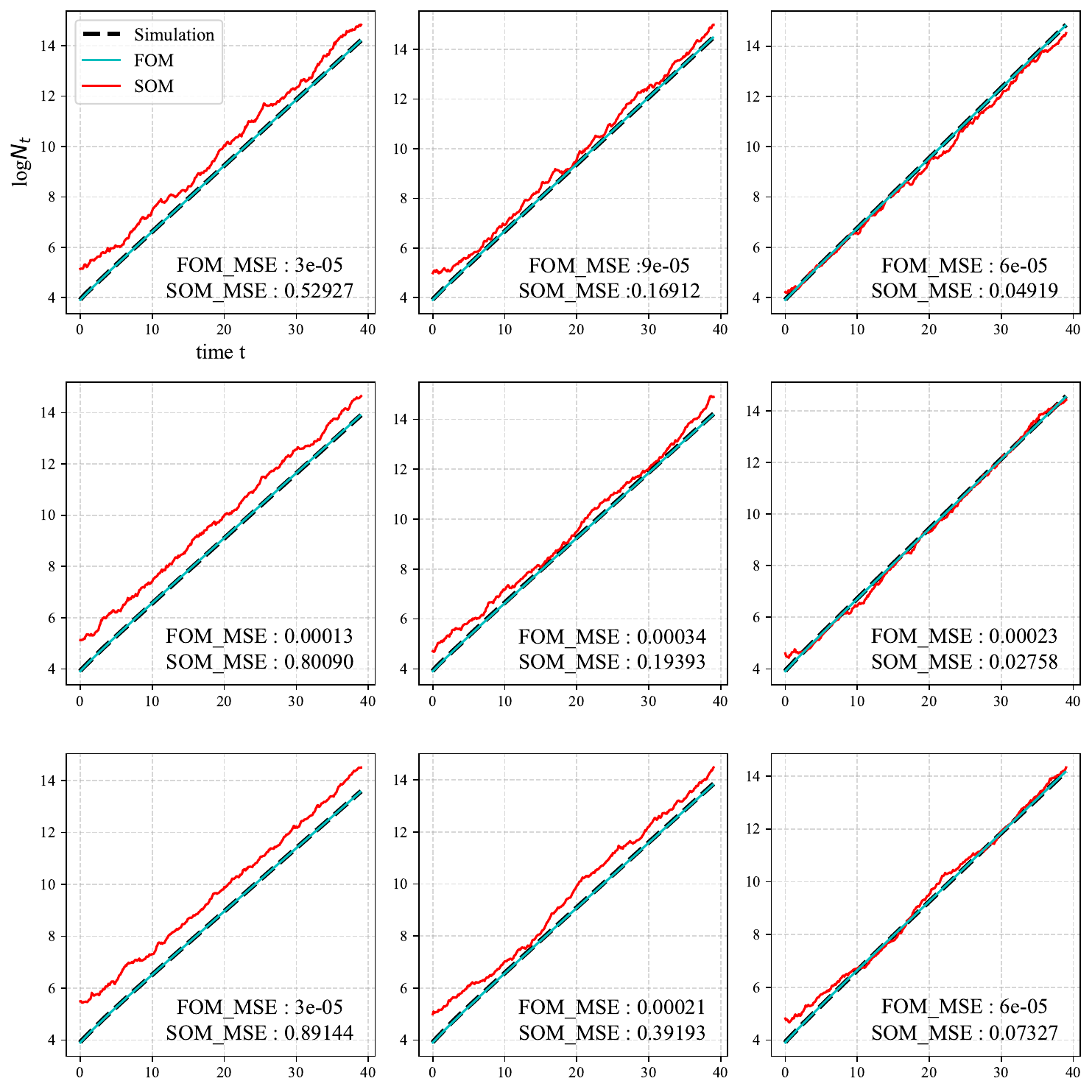}
\caption{$N_0=50$. Parameters:($r_1, r_2, p_1, \delta_2$) = (0.4, 0.2, 0.35, 0.25), the number of trajectories is 100.}
\label{N_0=50}
\end{figure}

\begin{figure}[H]
\centering
\includegraphics[width=0.8\textwidth]{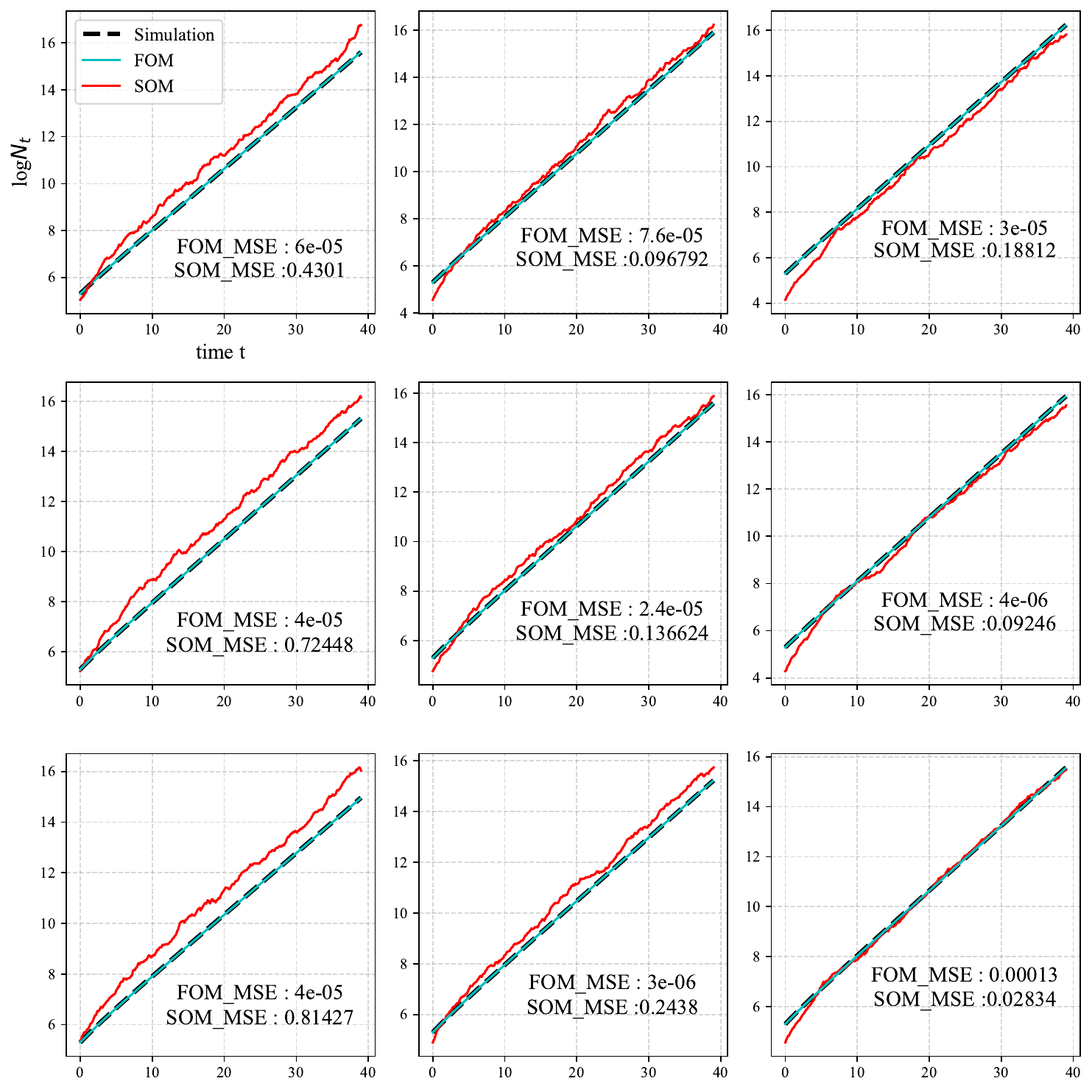}
\caption{$N_0=200$. Parameters:($r_1, r_2, p_1, \delta_2$) = (0.4, 0.2, 0.35, 0.25), the number of trajectories is 100.}
\label{N_0=200}
\end{figure}

Fig.\ref{N_0=50}, Fig. \ref{two_asym_total}, and Fig. \ref{N_0=200} respectively represent the performance of the FOM and SOM methods when the initial value $N_0$ is set to 50, 100, and 200. It shows that, although the predictive accuracy of both methods generally improves with increasing initial values, the overall variation is not pronounced.

\subsection{Different death rates}
\label{death rates}
In Model II, we considered a plasticity model without cell death, while in Model III, we incorporated cell death events. Comparing Fig.\ref{two_sym_total} and Fig. \ref{two_sym_death_total}, we find that the changes in the two results are not significant. we here further investigate the robustness of both methods to changes in the death rate. The parameters are configured as ($r_1, r_2, p_1, \delta_2$) = (0.4, 0.2, 0.35, 0.25).\\

\begin{figure}[H]
\centering
\includegraphics[width=0.8\textwidth]{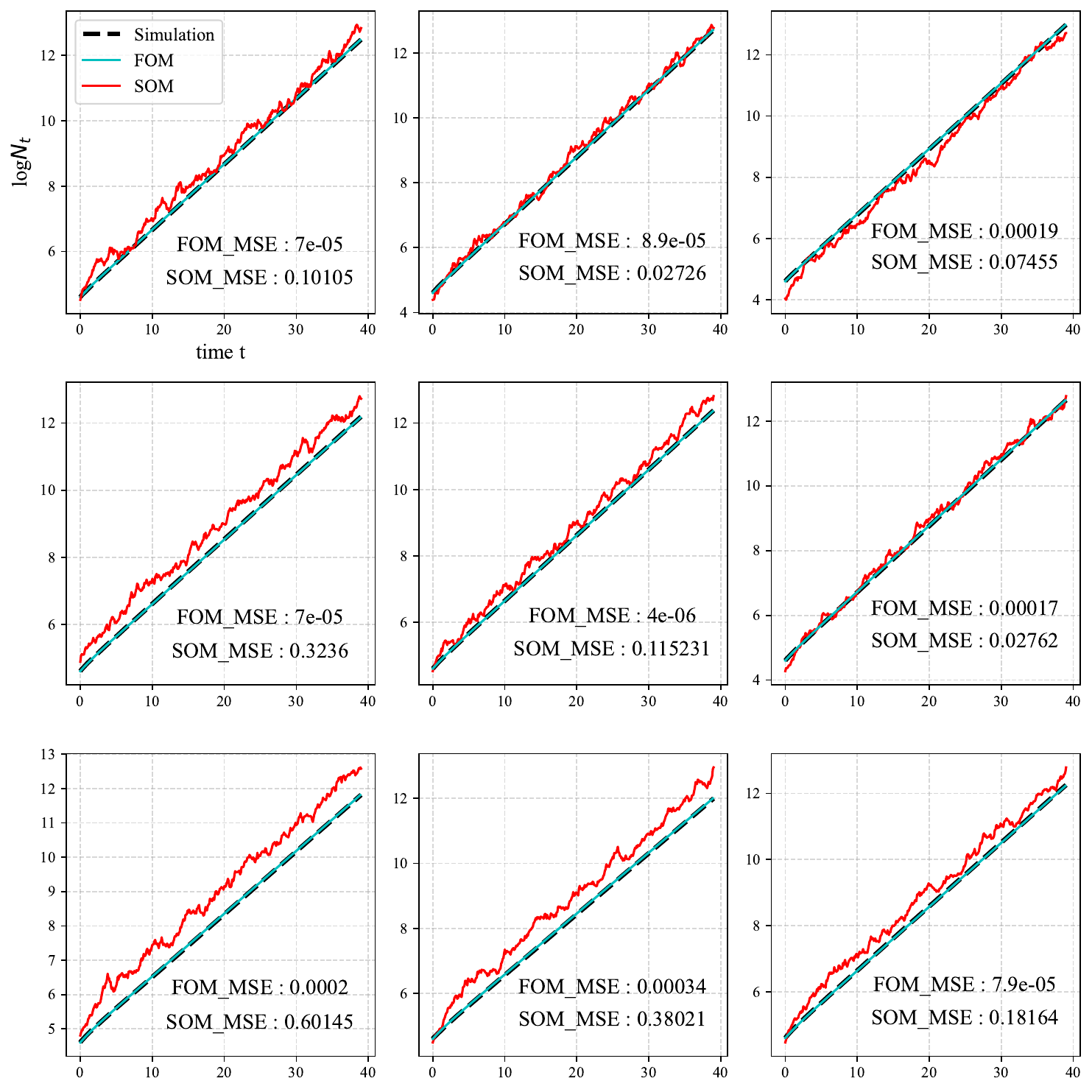}
\caption{llustration of the robustness of FOM and SOM in cell plasticity model III with $d$=0.05}
\label{d=0.05}
\end{figure}

By comparing Fig. \ref{two_sym_total}, Fig. \ref{two_sym_death_total}, and Fig. \ref{d=0.05}, it can be observed that the performance of both methods varies slightly across these three conditions, considering the absence of mortality events ($d$=0), low mortality rate ($d$=0.01), and high mortality rate ($d$=0.05), but the differences are not significant.

\subsection{Different levels of noise}
\label{noise}
During the process of experimental data collection, in addition to the stochasticity inherent in biological processes such as cell division that may cause internal noise, external noise can also arise due to factors such as measurement errors and environmental disturbances. We here augment the original simulated data with Gaussian noise, with variances of 0.01, 0.001, and 0.0001, respectively.\\
\begin{figure}[H]
    \centering
    \begin{subfigure}{0.24\textwidth}
        \centering
        \includegraphics[width=\textwidth]{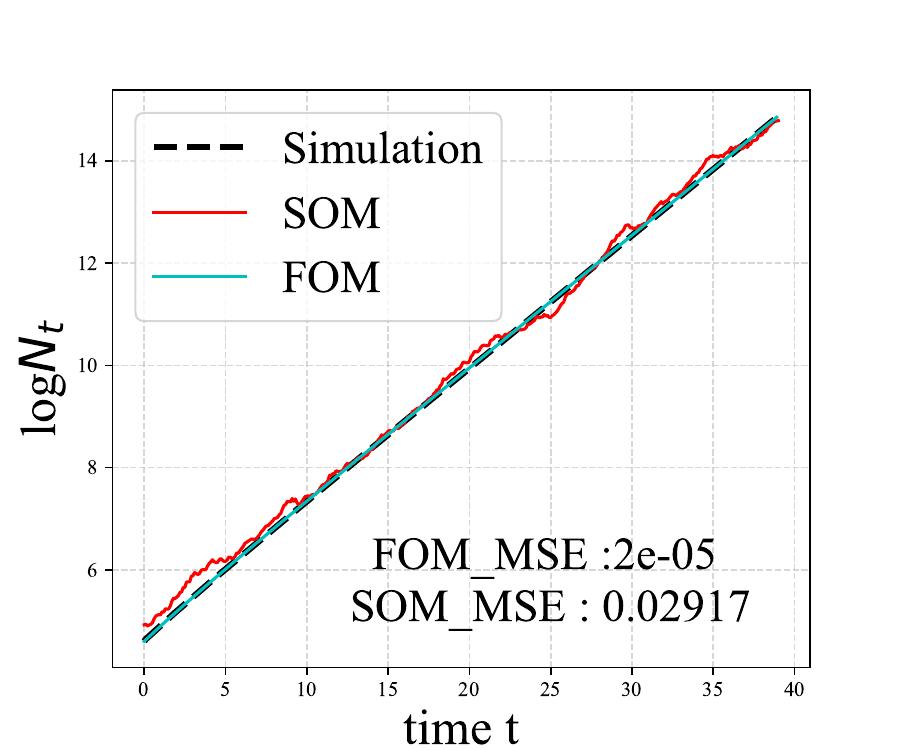}
        \caption{}
    \end{subfigure}%
    \begin{subfigure}{0.24\textwidth}
        \centering
        \includegraphics[width=\textwidth]{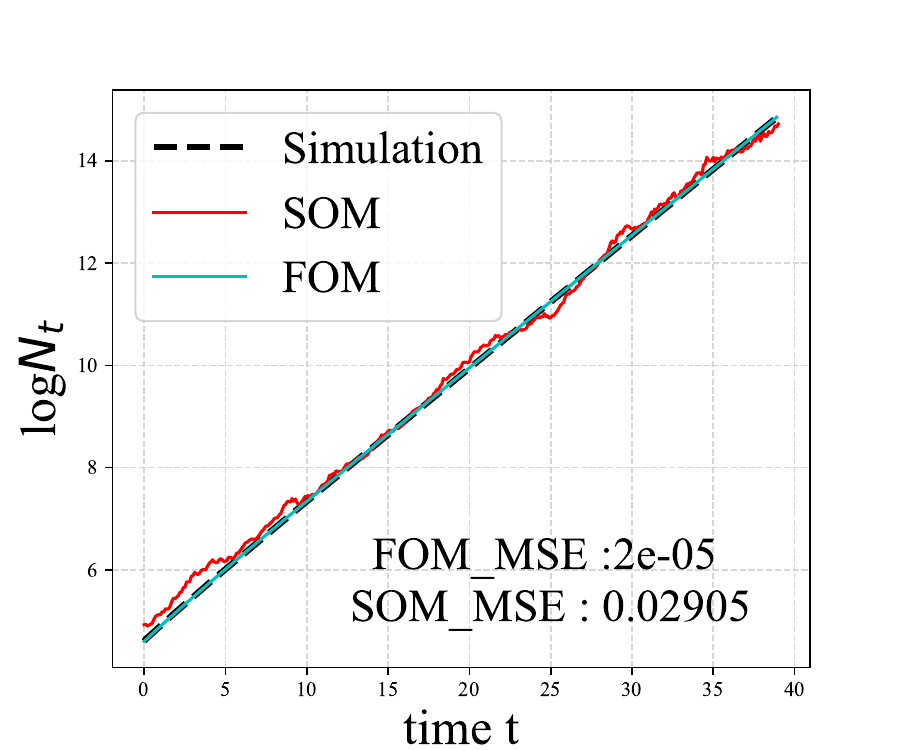}
        \caption{}
    \end{subfigure}%
    \begin{subfigure}{0.24\textwidth}
        \centering
        \includegraphics[width=\textwidth]{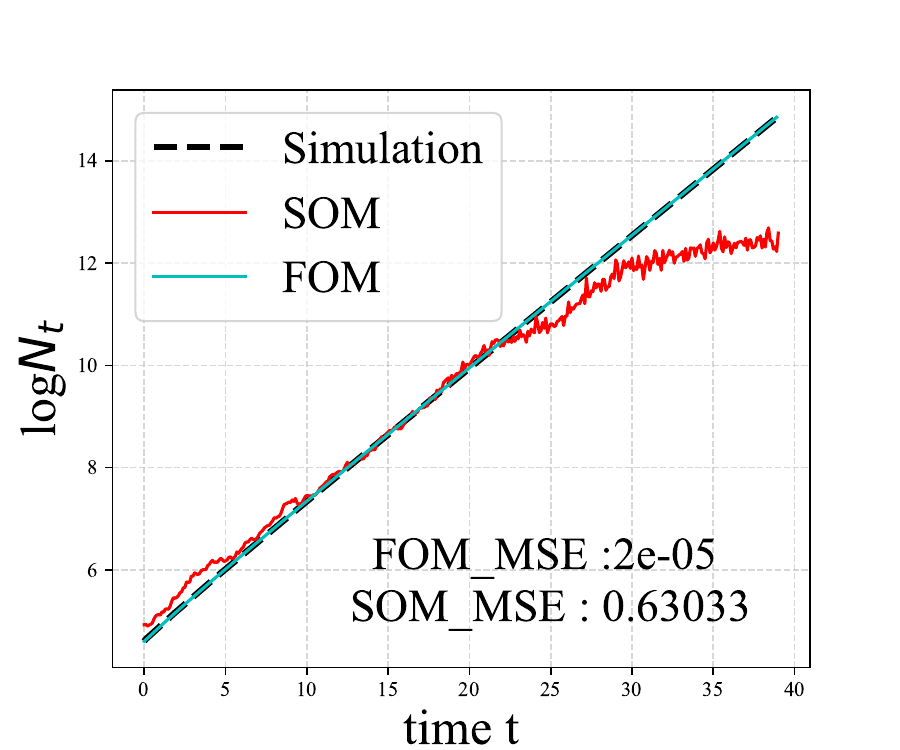}
        \caption{}
    \end{subfigure}%
    \begin{subfigure}{0.24\textwidth}
        \centering
        \includegraphics[width=\textwidth]{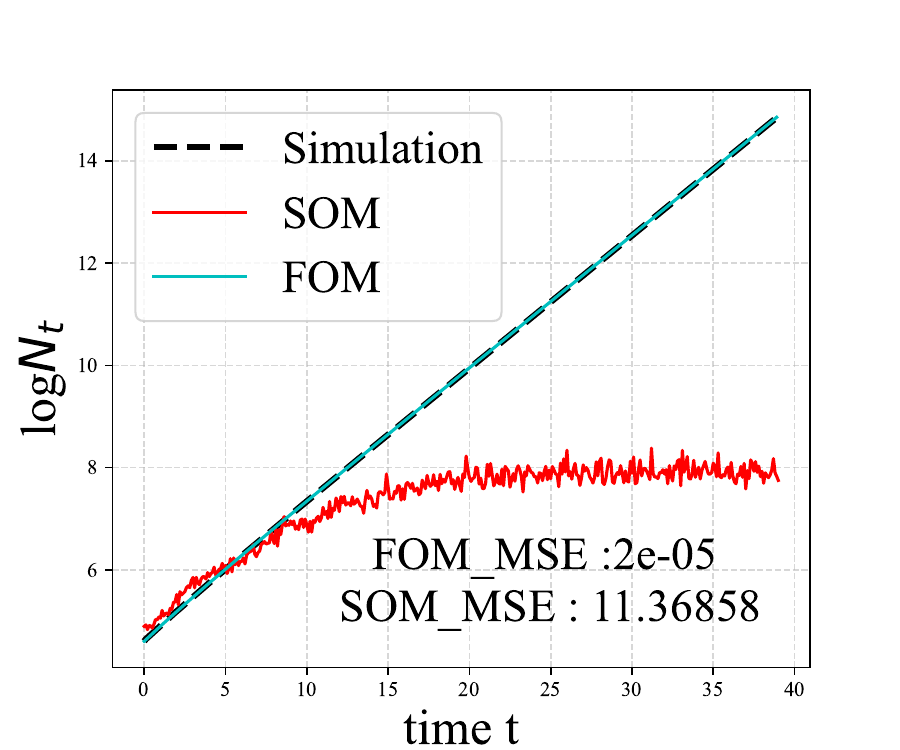}
        \caption{}
    \end{subfigure}
    \caption{The influence of different additional noise levels. Panel (a) represents conditions without additional noise interference, while panels (b), (c), and (d) are exposed to Gaussian noise with variances of 0.0001, 0.001, and 0.01, respectively. The parameters are configured as ($r_1, r_2, p_1, \delta_2$) = (0.4, 0.2, 0.35, 0.25), and the number of trajectories is 100.}
    \label{noise impact}
\end{figure}
Fig.\ref{noise impact} shows that the SOM method remains effective when the noise intensity is weak. However, as the noise intensity increases, the prediction by the SOM method starts to deviate and eventually becomes ineffective.

\subsection{Different sample frequencies}
\label{time interval}
In practical experiments, data is typically sampled with a certain frequency rather than recorded continuously in real-time. The problem of missing data occurs frequently. We here conducted sampling of simulated data at different time intervals. 

Tables \ref{tab: FOM} and \ref{tab: SOM} present the performance of the FOM and SOM methods under different time interval sampling conditions. We can see that, while the performance of the FOM method significantly deteriorates with decreasing sampling frequency, the SOM method is not as affected by the sampling frequency. We speculate that this might be attributed to the reduced sampling frequency affecting the approximation of numerical integral values in the FOM method, leading to a deterioration in its performance. In contrast, the SOM method appears to be unaffected by this issue, resulting in minimal changes in its accuracy. This intriguing finding suggests that the SOM method can still be utilized even in the presence of missing data.

\begin{table}[H]
\centering
\caption{The FOM method's effectiveness across different data sample frequency levels.}
\resizebox{1.0\columnwidth}{!}{ 
\begin{tabular}{|c|c|c|c|c|c|c|c|}
\hline
\diagbox{Parameters}{MSE}{Time intervals}  & $\Delta t$=0.1(n=400) & $\Delta t$=0.5(n=80)& $\Delta t$ =1(n=40)& $\Delta t$ =2(n=20) & $\Delta t$ =4(n=10)& $\Delta t$ =8(n=5)& $\Delta t$ =20(n=2)\\
\hline
($p_2,\delta_2$)=(0.25,0.35) &9e-05 &0.0001  & 0.00011 & 0.00018 &0.0006&0.000387&0.04056\\
\hline
($p_2,\delta_2$)=(0.25,0.45) & 1e-05 & 1e-05 & 1e-05 & 1e-05&0.00017&0.00177&0.02207 \\
\hline
($p_2,\delta_2$)=(0.35,0.35) & 0.00021 & 0.00022 & 0.00023 & 0.0003 &0.00057&0.00227&0.01995\\
\hline
($p_2,\delta_2$)=(0.35,0.45) & 0.00012 & 0.00012 & 0.00012 & 0.00015 &0.00027&0.00147&0.01243\\
\hline
\end{tabular}
}
\label{tab: FOM}
\end{table}

\begin{table}[H]
\centering
\caption{The SOM method's effectiveness across different data sample frequency levels.}
\resizebox{1.0\columnwidth}{!}{
\begin{tabular}{|c|c|c|c|c|c|c|c|}
\hline
\diagbox{Parameters}{MSE}{Time intervals}  & $\Delta t$=0.1(n=400) & $\Delta t$=0.5(n=80)& $\Delta t$ =1(n=40)& $\Delta t$ =2(n=20) & $\Delta t$ =4(n=10)& $\Delta t$ =8(n=5)& $\Delta t$ =20(n=2)\\
\hline
($p_2,\delta_2$)=(0.25,0.35) &0.17364 &0.17263  & 0.17202 & 0.1764&0.18341&0.1234&0.19024 \\
\hline
($p_2,\delta_2$)=(0.25,0.45) & 0.06795 & 0.06861 & 0.06781 & 0.06701 &0.0627&0.04471&0.0592\\
\hline
($p_2,\delta_2$)=(0.35,0.35) & 0.03435 & 0.03466 & 0.03422 & 0.03856&0.0304&0.0324&0.00838 \\
\hline
($p_2,\delta_2$)=(0.35,0.45) & 0.08028 & 0.07909 & 0.07973 & 0.07275 &0.06651&0.07481&0.05442\\
\hline
\end{tabular}
}
\label{tab: SOM}
\end{table}

\newpage
\bibliography{reference}

\end{document}